\documentclass[pra,aps,superscriptaddress,footinbib,twocolumn]{revtex4-2}
\usepackage{graphicx} 
\usepackage{amsmath}
\usepackage[colorlinks=true,citecolor=blue,linkcolor=magenta]{hyperref}
\usepackage[braket]{qcircuit}
\usepackage{appendix}
\usepackage{amsfonts}
\usepackage{amssymb}
\usepackage{amscd}
\usepackage{enumerate}
\usepackage{epsfig}
\usepackage{subfigure}
\usepackage{bm}
\usepackage{xcolor}
\usepackage{amsthm}
\usepackage{framed}
\usepackage{multirow}
\usepackage{mathrsfs,amssymb}
\usepackage{latexsym} 
\usepackage{physics}
\usepackage{epstopdf}
\usepackage[skins]{tcolorbox}
\newtcolorbox[auto counter]{tbox}[2][]{%
    enhanced, float=hbt, drop fuzzy shadow southeast,
    colback=white!5!white, colframe=white!50!black,
    width= .97\columnwidth,sharp corners, boxrule=0.8pt,
    title={Table \thetcbcounter: #2}, #1
}
\usepackage[linesnumbered,boxed,ruled,vlined]{algorithm2e}
\usepackage{bbm}
\usepackage{subfigure}

\newtheorem{theorem}{Theorem}[section]
\newcommand{\X}{\hat{\rho}}

\begin{document}
\title{Error mitigated shadow estimation based on virtual distillation}
\author{Ruyu Yang}
\affiliation{State Key Lab of Processors, Institute of Computing Technology,
Chinese Academy of Sciences, 100190, Beijing, China}
\affiliation{Graduate School of China Academy of Engineering Physics, Beijing 100193, China}

\author{Xiaoming Sun}
\affiliation{State Key Lab of Processors, Institute of Computing Technology,
Chinese Academy of Sciences, 100190, Beijing, China}

\author{Hongyi Zhou}
\email{
zhouhongyi@ict.ac.cn
}

\affiliation{State Key Lab of Processors, Institute of Computing Technology,
Chinese Academy of Sciences, 100190, Beijing, China}

\begin{abstract}

Shadow estimation is a method for deducing numerous properties of an unknown quantum state through a limited set of measurements, which suffers from noises in quantum devices. In this paper, we introduce an error-mitigated shadow estimation approach based on virtual distillation, tailored for applications in near-term quantum devices.
Our methodology leverages the qubit reset technique, thereby reducing the associated qubit overhead. Crucially, our approach ensures that the required qubit resources remain independent of the desired accuracy and avoid an exponential measurement overhead, marking a substantial advancement in practical applications. Furthermore, our technique accommodates a mixed Clifford and Pauli-type shadow, which can result in a reduction in the number of required measurements across various scenarios.
We also study the trade-off between circuit depth and measurement overhead quantitatively. 
Through numerical simulations, we substantiate the efficacy of our error mitigation method, establishing its utility in enhancing the robustness of shadow estimations on near-term quantum devices.
\end{abstract}
\maketitle

\section{Introduction}

Shadow estimation enables predictions of numerous properties of an unknown quantum state through a limited set of measurements, which was originated from Aaronson~\cite{aaronson2018shadow} and further developed by Huang et al.~\cite{huang2020predicting}. The core idea of shadow estimation is to construct the classical shadow, a classical representation of a quantum state that can be efficiently computed from random stabilizer measurements. The various applications include the estimation of entanglement~\cite{zhang2021experimental,PhysRevLett.125.200501,PhysRevLett.125.200502,PhysRevLett.129.260501,neven2021symmetry,li2021vsql}, fidelity~\cite{struchalin2021experimental}, entropy~\cite{sack2022avoiding}, and other applications~\cite{hadfield2022measurements,hadfield2021adaptive,garcia2021quantum,mcginley2022quantifying,huggins2022unbiasing,huang2022provably,notarnicola2023randomized,helsen2023shadow,wu2023overlapped,yang2023shadow,wan2023matchgate}. In this work, we focus on the shadow estimations on the linear $\Tr(O\rho)$ and nonlinear $\Tr(O_1\rho O_2\rho\cdots O_p\rho)$ functions of quantum states. The linear functions are the expectations of physical observables or nonphysical operators. While the general nonlinear functions also play an important role in quantum computation tasks. In the variational preparation of non-equilibrium steady states ~\cite{liu2021variational,zhou2023hybrid}, the cost function, which is the Frobenius norm of the Lindblad equation, has a form of $\Tr(O_1\rho O_2\rho)$. On the other hand, a more general form, $\Tr(O_1\rho O_2\rho\cdots O_p\rho)$, can also be used for calculating the quantum Fisher information~\cite{rath2021quantum}.

Current quantum devices are prone to decoherence due to environmental interactions~\cite{chirolli2008decoherence,devoret2004superconducting,bruzewicz2019trapped,preskill2018quantum}, leading to errors in quantum computing and shadow estimation. 
These errors are categorized as state preparation error, measurement error, and gate error based on the location of occurrence. As quantum computers scale up in size and complexity, the likelihood of errors also accumulates, especially for the gate error. While state preparation and measurement (SPAM) errors in shadow estimation have been addressed in previous works~\cite{chen2021robust, koh2022classical,bu2024classical,wu2023error}, the gate errors remain a formidable challenge. Recent attempts, such as a classical shadow based on Probabilistic Error Cancellation~\cite{jnane2023quantum}, require prior characterization of error models, presenting difficulties for large systems.

Another approach to address gate errors in shadow estimation involves estimating nonlinear functions of a noisy quantum state $\Tr(\rho_e^M O)/\Tr(\rho_e^M)$ instead of directly estimating $\Tr(\rho_e O)$. This is inspired by an error mitigation method known as virtual distillation (VD) \cite{huggins2021virtual, koczor2021exponential,koczor2021dominant}, which only assumes a stochastic error model and offers exponential error suppression. In fact, the coherent noise can also be transformed into stochastic noise by using the Pauli twirling technique, making VD applicable~\cite{cai2020mitigating,wallman2016noise,hashim2020randomized}. The fundamental concept involves performing collective measurements on $M$ copies of a noisy quantum state $\rho_e$ to calculate expectation values relative to the state, $\rho_e^M/\mathrm{tr}(\rho_e^M)$. Intuitively, the relative weights of non-dominant eigenvectors are exponentially suppressed as $M$ increases. The same idea can be applied to nonlinear functions $\Tr(O_1\rho_e^{M_1}O_2\rho_e^{M_2}\cdots O_p\rho_e^{M_p})/\Tr(\rho_e^{M_1})\Tr(\rho_e^{M_2})\cdots\Tr(\rho_e^{M_p})$.
In the original VD approach, the implementation becomes challenging when $M$ is large, as it necessitates the preparation of multiple copies of a noisy quantum system~\cite{bovino2005direct,horodecki2003measuring,ekert2002direct}. The shadow method allows the estimations for large $M$ without the need of actually preparing the multiple copies~\cite{huang2020predicting}. Nevertheless, this advantage comes at the cost of an exponential measurement overhead, which also becomes infeasible for large $M$. A hybrid framework, combining the generalized swap test and shadow estimation, has been proposed to reduce measurement overhead \cite{zhou2022hybrid}, which is still left exponential in $M$. 
In Table~\ref{Table: related work}, a comparison of these methods is presented, considering qubit resource requirements, circuit depth, and measurement overhead.

In this paper, we propose a novel shadow estimation method that mitigates the gate noise based on virtual distillation. This method is feasible in near-term quantum devices even for large $M$.
We use the technique of qubit reset to reduce the qubit overhead. Our approach ensures that the qubit resource requirements are independent of the accuracy considerations and at the meantime avoid the exponential measurement overhead. Moreover, our method allows for a mixed-type shadow, instead of using only Clifford- or Pauli-type shadow as in the conventional methods. This feature enables us to reduce the number of measurements with prior knowledge of the observables, compared with using a single type of shadow. We also investigate the calculation of the nonlinear functions with a large $M$ via shallower circuits with a depth of $M/a-1$ and analyze the corresponding variance. In particular, for the Clifford-type shadow and $d$-dimensional Hilbert space, the variance is $\sim\mathcal{O}(d/N)^a$, where $N$ denotes the number of shots. We evaluate the effectiveness of our error-mitigated shadow estimation method using numerical simulations. Through these advancements, our research strives to enhance the efficiency and feasibility of shadow estimation in quantum computing, paving the way for more robust and scalable quantum computations.

This paper is organized as follows. We revisit classical shadow and VD in Sec.~\ref{sec: pre}. We present our algorithm in Sec.~\ref{sec: alg}. We analyze the variance of the shadow estimations in Sec.~\ref{sec: scaling}. We show the numerical simulations in Sec.~\ref{sec: num}. We summarize this work in Sec.~\ref{sec: Con}.
\begin{table*}[t]
\begin{tabular}{|l|l|l|l|}
\hline
                     & qubits required & depth of circuit   & measurements overhead \\ \hline
Shadow with PEC~\cite{jnane2023quantum} &$n$ &constant &$\mathcal{O}(\exp\{ n \}\log(F))$  \\ \hline
Virtual distillation~\cite{huggins2021virtual, koczor2021exponential} & $nM+1$   & $\mathcal{O}(M)$     & $\mathcal{O}(F)$             \\ \hline
Hybrid framework \cite{zhou2022hybrid}              & $nt + 1$     & $\mathcal{O}(t)$     & $\mathcal{O}(\exp{n\lceil M/t\rceil}\log(F))$        \\ \hline
Shadow distillation \cite{seif2023shadow}             & $n$      & constant & $\mathcal{O}(\exp{nM}\log(F))$       \\ \hline
This work            & $2n+1$   & $\mathcal{O}(M)$     & $\mathcal{O}(M\log(F))$        \\ \hline
\end{tabular}
\label{Table: related work}
\caption{{Comparison of algorithms of original virtual distillation and error-mitigated shadow estimation. Here $n$ is the number of qubits of $\rho$, $M$ is the order of virtual distillation, $F$ is the number of observables, and $t=1,2,\cdots,M-1$ is the number of copies of $\rho$ in each group in the hybrid framework \cite{zhou2022hybrid}. Strictly speaking, the measurement overhead of the PEC (probabilistic error cancellation) method increases exponentially with the noisy quantum gate $n_e$, which usually tends to increase polynomially with the number of qubits $n$. }}
\end{table*}
\section{Preliminaries}

\label{sec: pre}

The steps for constructing a classical shadow are as follows.
\begin{itemize}
    \item Randomly insert global Clifford gates $C_i$ for Clifford-type shadow or local Clifford gate for Pauli-type shadow before measurement. Here the label $i$ represents the $i$-th run to perform this step. It is possible that $C_i = C_{i^\prime}$ for some $i\neq i^\prime$. This step is performed on a quantum computer.
    It is important to note that the type of Clifford gates inserted determines the type of classical shadow that is constructed.
    \item Measure the final state in computational basis $Z_i$ and obtain the result $| {z}_i \rangle$ where ${z}_i$ is a binary vector. This step is performed on a quantum computer. The binary vector $| {z}_i \rangle$ denotes the outcome of the measurement.\label{setp: CZ}
    \item Calculate the density matrix $\hat{\rho}_i=C^{\dagger}_i |{z}_i\rangle\langle {z}_i|C_i$. This step and all subsequent steps are performed on a classical computer. It is important to note that $\hat{\rho}_i$ is Hermitian, i.e., $\hat{\rho}_i = \hat{\rho}^{\dagger}_i$. In particular, when $C_i$ is the tensor product of single-qubit Clifford gate, the density matrix $\hat{\rho}_i$ can be expressed as $\hat{\rho}_i = \bigotimes_j \hat{\rho}_{i}^j$ where $\hat{\rho}_{i}^j$ is the density matrix defined on $j$-th qubit.
    \item Apply the linear map \begin{equation}
        \tilde{\rho}_i = \mathcal{M}_C(\hat{\rho}_i) = (2^n + 1)\hat{\rho}_i - I_{2^n},
    \end{equation}
    for Clifford-type shadow, or 
    \begin{equation}
        \tilde{\rho}_i = \mathcal{M}_P(\hat{\rho}_i) = \bigotimes_{j}(3\hat{\rho}_{i}^j - I_2),
    \end{equation}
    for Pauli-type shadow, where $I_x$ denotes a $x$ by $x$ identity matrix. 
    The linear maps $\mathcal{M}_C$ and $\mathcal{M}_P$ transform the density matrix into the shadow representation for Clifford-type and Pauli-type shadows, respectively.
    \item Repeat the steps above for $N$ times and calculate the average $\tilde{\rho} = \frac{1}{N}\sum_i\tilde{\rho}_i$ on the classical computer, where $\tilde{\rho}_i$ is the classical shadow obtained from single-shot. The estimator of the observable $O$ can be obtained from $\Tr(O\Tilde{\rho})$.

\end{itemize}

The shadow $\tilde{\rho}$ constructed by the above steps is still affected by noises in the quantum circuit. To mitigate this noise, VD can be applied by transforming linear observations into nonlinear ones. As a result, an error-mitigated shadow can be obtained. As shown in Fig.~\ref{fig:original}, the final state before randomized measurement is
\begin{align}
    \rho_f & = \frac{1}{2}|0\rangle\langle0| \otimes \rho^{\otimes M} + \frac{1}{2}|0\rangle\langle1|\otimes\rho^{\otimes M}S^{\dagger}_M \nonumber
    \\ &+ \frac{1}{2}|1\rangle\langle0|\otimes S_M\rho^{\otimes M} + \frac{1}{2}|1\rangle\langle1|\otimes S_M\rho^{\otimes M}S_M^{\dagger},
    \label{eq: final state of swap test}
\end{align}
where $M$ is the order of VD. The general swap operator $S_M$ is defined as
\begin{align}
    S_M |\phi_1\rangle|\phi_2\rangle\dots|\phi_M\rangle = |\phi_M\rangle|\phi_1\rangle\dots|\phi_{M-1}\rangle,
\end{align}
where $|\phi_i\rangle$ ($i =1,2,\dots,M$) are arbitrary $n$-qubit states. According to the properties of classical shadow, we can assert that
\begin{align}
    \mathbf{E}_i \left[\Tr(\Tilde{O} \tilde{\rho}_{i})\right] = \Tr(\Tilde{O}\rho_f),\label{eq: average of classical shadow}
\end{align}
where $\tilde{\rho}_{i}$ is the single-shot classical shadow obtained by running the quantum circuit for the $i$-th round and $\Tilde{O}$ is an arbitrary linear operator. The variance of the estimation with finite shots depends on the specific form of $\Tilde{O}$ and the selection method of the Clifford gate $C_i$. It can be proven that if $\Tilde{O} = f_{X_0} + if_{Y_0}$, where $f_A = A\otimes O_1\otimes O_2\otimes\dots \otimes O_M$, the right hand side of Eq.~\eqref{eq: average of classical shadow} is equal to $\Tr(O_1\rho_e O_2\rho_e\dots O_M\rho_e)$. 
It is straightforward to find that
\begin{equation}
\begin{aligned}
    &\Tr(X_0\otimes O_1\otimes\dots\otimes O_M\rho_f)\\  =& \frac{1}{2}\Tr(\rho_e^{\otimes M}S_M^{\dagger}O_1\otimes\dots\otimes O_M) \\
    &+ \frac{1}{2}\Tr(S_M\rho_e^{\otimes M}O_1\otimes\dots\otimes O_M) \\
    =&
    \mathbf{Re}[\Tr(S_M\rho_e^{\otimes M}O_1\otimes\dots\otimes O_M)],
\end{aligned}
\end{equation}
and
\begin{equation}
\begin{aligned}
    &\Tr(Y_0\otimes O_1\otimes\dots\otimes O_M\rho_f) \\  = &-\frac{i}{2}\Tr(\rho_e^{\otimes M}S_M^{\dagger}O_1\otimes\dots\otimes O_M) \\
    &+ \frac{i}{2}\Tr(S_M\rho_e^{\otimes M}O_1\otimes\dots\otimes O_M) \\
    =&
\mathbf{Im}[\Tr(S_M\rho_e^{\otimes M}O_1\otimes\dots\otimes O_M)].
\end{aligned}
\end{equation}
Combining the above two equations, we obtain that 
\begin{equation}
\begin{aligned}
&\Tr(\Tilde{O}\rho_f)\\= &\mathbf{Re}[\Tr(S_M\rho_e^{\otimes M}O_1\otimes \dots \otimes O_M)]  \\
& + i\mathbf{Im}[\Tr(S_M\rho_e^{\otimes M}O_1\otimes \dots \otimes O_M)]
    \\ = &\Tr(S_M\rho_e^{\otimes M}O_1\otimes \dots \otimes O_M)\\ = &
    \sum_{j_1,\dots,j_M} p_{j_1}\cdots p_{j_M}\Tr\left[S_M (|\phi_{j_1}\rangle\langle\phi_{j_1}|\otimes \dots \otimes |\phi_{j_M}\rangle\langle\phi_{j_M}| )  \right.\\
   & \quad \left.\times O_1\otimes O_{2}\otimes \dots \otimes O_M \right]\\=& \sum_{j_1,\dots,j_M}p_{j_1}\cdots p_{j_M}\Tr\left[(|\phi_{j_M}\rangle\langle\phi_{j_1}|\otimes \dots \otimes |\phi_{j_{M-1}}\rangle\langle\phi_{j_M}|) \right. \\
   & \quad \left. \times O_1\otimes O_{2} \otimes \dots \otimes O_M\right]\\
=&\sum_{j_1,\dots,j_M}p_{j_1}\cdots p_{j_M}\langle\phi_{j_1}|O_1|\phi_{j_M}\rangle\dots\langle\phi_{j_M}|O_M|\phi_{j_{M-1}}\rangle\\
=&\Tr(O_1\rho_e O_{2}\rho_e\cdots O_M\rho_e)\label{eq: aimed observations },
\end{aligned}
\end{equation} 
where $\rho_e = \sum_{j}p_j|\phi_{j}\rangle\langle\phi_{j}|$ ($\sum_j p_j = 1$), and $\langle\phi_{j}|\phi_{j^{\prime}}\rangle$ needs not to be zero for $j\neq j^\prime$.
However, this method suffers from the drawback of requiring large qubit resources. For instance, computing an $M$-order function of a $n$-qubit state requires at least $nM+1$ qubits in the experiment. This poses a significant challenge in the NISQ era, where the number of accessible qubits is still limited. Hence, it is crucial to devise algorithms that are more efficient in terms of qubit consumption.
\section{ALGORITHM}
\label{sec: alg}
\begin{figure*}
    \centerline{
\Qcircuit @C = 1em @R = 2em{
\lstick{ \ket{0}} & \qw   & \gate{H}  &\qw &\ctrl{2} &\qw &\qw  &\ctrl{3}  &\qw &\qw &\ctrl{4} &\qw &\qw &\cdots&~&\qw &\qw &\ctrl{5} &\qw &\qw &\multigate{5}{\mathcal{C}_i}&\qw &\qw &\meter  \\    
\lstick{\ket{0}^{\otimes n}} & {/}\qw &\gate{U}   &\qw &\qswap &\qw &\qw  &\qswap   &\qw &\qw &\qswap &\qw &\qw &\cdots&~ &\qw &\qw &\qswap&\qw  &\qw&\ghost{\mathcal{C}_i} &\qw &\qw &\meter\\    
\lstick{\ket{0}^{\otimes n}} & {/}\qw &\gate{U}   &\qw &\qswap &\qw &\qw  &\qw      &\qw &\qw &\qw &\qw &\qw &\cdots &~&\qw &\qw  &\qw&\qw &\qw &\ghost{\mathcal{C}_i} &\qw &\qw &\meter\\    
\lstick{\ket{0}^{\otimes n}} & {/}\qw &\gate{U}   &\qw &\qw
&\qw &\qw  &\qswap   &\qw &\qw &\qw &\qw &\qw &\cdots&~ &\qw &\qw &\qw&\qw &\qw &\ghost{\mathcal{C}_i} &\qw &\qw &\meter\\   
\lstick{\ket{0}^{\otimes n}} & {/}\qw &\gate{U}   &\qw &\qw
&\qw  &\qw &\qw &\qw &\qw  &\qswap &\qw  &\qw&\cdots&~&\qw &\qw &\qw &\qw &\qw  &\ghost{\mathcal{C}_i} &\qw &\qw &\meter\\  
\lstick{\ket{0}^{\otimes n}} & {/}\qw &\gate{U}   &\qw &\qw
&\qw  &\qw &\qw &\qw &\qw  &\qw &\qw  &\qw &\cdots&~&\qw &\qw &\qswap &\qw &\qw &\ghost{\mathcal{C}_i} &\qw &\qw &\meter\\  
}
}
    \caption{Swap test circuit consists of a Hadamard test and a cyclic permutation operation. $H$ is the Hadamard gate. $C_i$ is the Clifford gate applied during the $i$-th run of the circuit. The total number of qubits is $nM+1$. The circuit depth scales as $\sim \mathcal{O}(M)$. Before the final Pauli-$Z$ measurements, a random global Clifford gate is applied. The unitary operation $U$ is a priori for the preparation of the target quantum state $\rho$. In this circuit, the cyclic permutation operator is implemented as a right shift. }
    \label{fig:original}
\end{figure*}
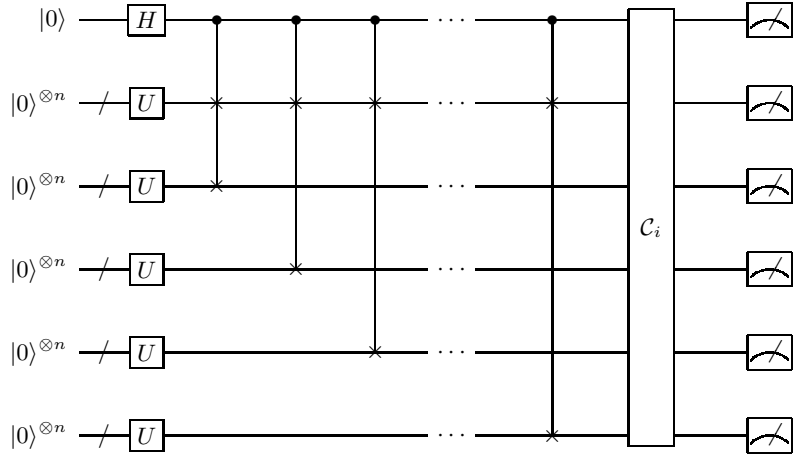

\begin{figure*}
    \centerline{
\Qcircuit @C = 0.8em @R = 2em{
\lstick{\ket{0}} &\qw &\gate{H} &\qw &\ctrl{2} &\qw &\qw&\qw&\qw&\qw&\qw&\qw&\qw&\qw&\qw&\qw&\qw&\ctrl{2}&\qw&\qw&\qw&\qw&&\cdots&~&~&&\qw&\qw&\qw&\qw& \qw&\ctrl{2}&\qw&\gate{C_{0,i}}&\qw&\meter\\
\lstick{\ket{0}^{\otimes n}} &{/}\qw &\gate{U} &\qw &\qswap &\qw &\qw&\qw&\qw&\qw&\qw&\qw&\qw&\qw&\qw&\qw&\qw&\qswap&\qw&\qw&\qw&\qw&&\cdots&~&~&&\qw&\qw&\qw&\qw&\qw&\qswap&\qw&\gate{C_{1,i}^A}&\qw&\meter\\
\lstick{\ket{0}^{\otimes n}} &{/}\qw &\gate{U} &\qw &\qswap &\qw &\gate{C_{M,i}^B} &\qw &\meter &&&&&\lstick{\ket{0}^{\otimes n}}&{/}\qw&\gate{U}&\qw&\qswap&\qw &\gate{C_{M-1,i}^B} &\qw &\meter&&\cdots&~&~ &&&\lstick{\ket{0}^{\otimes n}}&{/}\qw&\gate{U}&\qw&\qswap&\qw&\gate{C_{2,i}^B}&\qw&\meter\\
}
}
    \caption{Quantum circuit with qubits reset. The number of qubits is $2n+1$, irrelevant with the order $M$. The circuit consists of an ancillary qubit, Register $A$, and Register $B$. Register $B$ will be reset to state $|0\rangle^{\otimes n}$ for $M-2$ times. $H$ is the Hadamard gate. $U$ is the unitary gate such that $\rho = U|0\rangle\langle 0|U^{\dagger}. $ $C_{0,i}$ and $C_{1,i}^A$ are the Clifford gates applied on the ancilla qubit and register $A$, i.e., the first $n$-qubit subsystem, during the $i$-th run of the algorithm, respectively. $C_{j,i}^B$ is the Clifford gate applied on the $j$-th $n$-qubit subsystem in register $B$ ($j\in\{2,3,\dotsc M-1\}$). Global or local Clifford gates are applied randomly before each qubit reset and the final Pauli-$Z$ measurements.}
    \label{fig: qubit efficient}
\end{figure*}
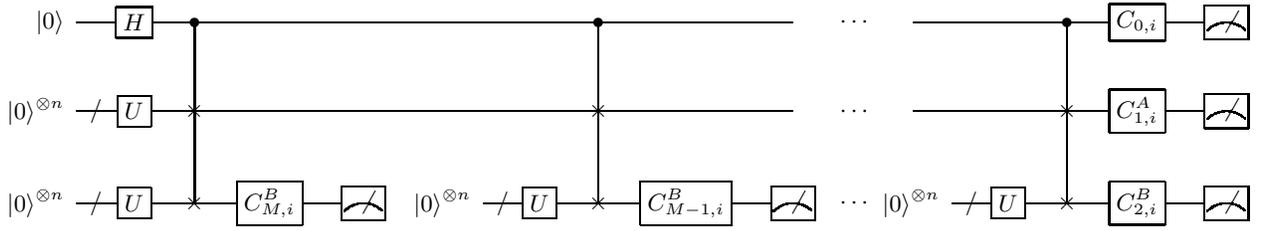

In this section, we present a qubit-efficient error-mitigated shadow estimation method that uses the qubit reset operation~\cite{yirka2021qubit}. This method only needs $2n+1$ qubits, irrespective of $M$. The quantum circuit is given in Fig.~\ref{fig: qubit efficient}.
The key idea of this method is to exploit the fact that most copies of $\rho_e$ pass through the controlled-SWAP gate only once, as illustrated in Fig.~\ref{fig:original}. 
We can easily check that the circuits in Fig.~\ref{fig:original} and Fig.~\ref{fig: qubit efficient} are equivalent if we ignore the random Clifford gates before the measurement. An important observation is that this method reduces the number of qubits without increasing the circuit depth, i,e, the requirement of the Toffoli gate is the same as the original swap test method.
However, when $O_1, O_2, \dots, O_M$ are global operators defined on the Hilbert space of $\rho_e$, reducing qubits will incur corresponding costs. Since we cannot apply a global Clifford gate on the Hilbert space of $\rho_e^{\otimes M}$, the variance of the estimation will grow exponentially with the number of global operators. We show the specific result in Theorem~\ref{thm-1}. In this case, the classical shadow has a form between Pauli-type and complete Clifford-type shadows. 


We define the linear transformation ${\mathcal{\widetilde{M}}_C}$ as
\begin{align}
    &   \mathcal{\widetilde{M}}_C(R^0\otimes \Tilde{R}^{(1)}\otimes  \dots \otimes \Tilde{R}^{(M)})\nonumber\\
     =& \mathcal{M}_P(R^0) \otimes \mathcal{M}_{C}(\Tilde{R}^{(1)}) \otimes \dots  \otimes \mathcal{M}_{C}(\Tilde{R}^{(M)})\nonumber\\
     = &(3R^0 - I_2) \otimes ((2^n+1)\Tilde{R}^{(1)} - I_{2^n}) \otimes \dots \nonumber\\
    & \otimes ((2^n+1)\Tilde{R}^{(M)} - I_{2^n}),
\end{align}
where $R^0$ is the Hermitian operator defined on a single qubit with unit trace and $\Tilde{R}^{(j)}$ ($j=1,2,\dotsc, M$) is the Hermitian operator defined on the $j$-th $n$-qubit subsystem with unit trace. The $n$-qubit subsystems are labeled by $1,\dotsc,M$ as shown in Fig.~\ref{fig: qubit efficient}. For the Pauli-type shadow model where the random gates $C_{j,i}^B$ and $C_{1,i}^A$ are the tensor products of the local Clifford gates, the linear map $\mathcal{\widetilde{M}}_P$ can be expressed as
\begin{align}
    &\quad  \mathcal{\widetilde{M}}_P(R^0\otimes R^1\otimes  \dots \otimes R^{nM}) \nonumber\\
    & = \mathcal{M}_P(R^0) \otimes \mathcal{M}_P(R^1) \otimes \dots  \otimes \mathcal{M}_P(R^{nM})\nonumber\\
    & = (3R^0 - I_2) \otimes (3R^1 - I_2) \otimes \dots \otimes (3R^{nM} - I_2),
\end{align}
where $R^q$ is a Hermitian operator with unit trace defined on the $q$-th qubit for $q=0, 1,2,\dots,nM$.
It is noteworthy that when some observables $O_1,\dotsc,O_{j_0}$ are local and some other observables $O_{{j_0}+1},\dotsc,O_M$ are global, we can choose a more flexible strategy between the two approaches described above. We choose the corresponding random Clifford gate to be either the tensor product of the local Clifford gates or the global Clifford gate, depending on whether the observable $O$ is local or global. Consequently, the linear map $\mathcal{\widetilde{M}}_{\mathrm{mix}}$ can be written as
\begin{align}
&\mathcal{\widetilde{M}}_{\mathrm{mix}}(R^0\otimes R^1\otimes  \dots \otimes R^{j_0n} \otimes \tilde{R}^{(j_0+1)}\otimes\cdots\otimes\Tilde{R}^{(M)}) \nonumber\\ = &(3R^0 - I_2) \otimes (3R^1 - I_2) \otimes (3R^2 - I_2) \otimes \dots \nonumber\\ &\otimes (3R^{j_0n} - I_2)
     \otimes ((2^n+1)\Tilde{R}^{(j_0+1)} - I_{2^n}) \otimes \dots\nonumber\\ &\otimes ((2^n+1)\Tilde{R}^{(M)} - I_{2^n}).
\end{align}
We summarize the overall algorithm in Algorithm~\ref{alg:QESE}.
\begin{algorithm}[htb]\caption{Qubit-efficient shadow estimation}\label{alg:QESE}
\KwIn{Order $M$, number of measurements $N$.}
\textbf{For} $i=1:1:N$   \\
\quad  Initialize $2n+1$ qubits to $|0\rangle^{\otimes 2n+1}$, numbered $0,1,.....,2n$. Qubit $0$ is the ancillary qubit, qubits $1,2,... n$ is denoted as register $A$, and qubits $n+1, n+2,.....,2n$ are denoted as register $B$. \\

\quad Apply a Hadamard gate on the ancillary qubit. Apply the unitary $U$ on registers $A$ and $B$, respectively. \\

\quad Apply a controlled-SWAP gate where the ancillary qubit is the control qubit. The role of the swap operation is to exchange the quantum states of register $A$ and register $B$. \\

\quad A Clifford gate $C_{M, i}^B$ defined on $n$ qubits is randomly selected to act on register $B$. Then take a Pauli-$Z$ measurement and get the measurement result $|z_{M,i}^B\rangle$. \\

\quad \textbf{For} $j = M-1:1:2$ \\
\quad\quad  Reset register $B$ to $|0\rangle^{\otimes n}$. \\
\quad\quad  Apply the unitary $U$ on register $B$. \\
\quad\quad Apply a controlled-SWAP gate between the ancillary qubit and two registers. \\
\quad\quad Apply a random Clifford gate $C_{j, i}^B$ on register $B$. Then perform Pauli-$Z$ measurements and obtain the measurement results $|z_{j,i}^B\rangle$.\\
\quad \textbf{end For} \\
\quad Apply a random Clifford gate $C_{1, i}^A$ on register $A$. Perform Pauli-$Z$ measurement and obtain the measurement result $|z_{1, i}^A\rangle$. \\
\quad Apply a random single-qubit Clifford gate $C_0$ on the ancillary qubit. Then a Pauli-$Z$ measurement is taken to obtain the result $|z_{0, i}\rangle$. \\

\quad Calcuate the single-shot shadow $\Tilde{\rho}_i = \mathcal{\widetilde{M}}(C_i|z_i\rangle\langle z_i|C_i^{\dagger})$, where 
\begin{equation}
    C_i = C_{M,i}^B\otimes C_{M-1,i}^B\otimes \dots \otimes C_{2,i}^B\otimes C_{1,i}^A\otimes C_{0,i},
\end{equation}
and 
{
\begin{equation}
|z_i\rangle = |z_{M,i}^B\rangle \otimes |z_{M-1,i}^B\rangle\otimes \dots \otimes |z_{2,i}^B\rangle\otimes |z_{1,i}^A\rangle\otimes |z_{0,i}\rangle. 
\end{equation}
}
The linear map $\mathcal{\widetilde{M}}$ can be $\mathcal{\widetilde{M}}_P$, $\mathcal{\widetilde{M}}_C$ or $\mathcal{\widetilde{M}}_{\mathrm{mix}}$.\\
\textbf{end For} \\
Calculate $\tilde{\rho} = \frac{1}{N}\sum_i\tilde{\rho}_i$.
\end{algorithm}

Note that the specific forms of $C^A_{1,i}$ and $C^B_{j,i}$ are not given. They can be either global Clifford gates or tensor products of single-qubit Clifford gates. $C_{0,i}$ is always a single-qubit Clifford gate because the measurements on the ancilla qubit are always $X_0$ or $Y_0$, which are both single-qubit Pauli operators.


In the NISQ era, both the number of qubits and the circuit depth are constrained. Quantum circuit errors often increase exponentially with circuit depth.  Hence, reducing the circuit depth of an algorithm will make it more compatible with NISQ machines. Even halving the circuit depth can result in a square root enhancement in the quantum circuit fidelity.
We propose a method to compute $M$-order quantum state functions with shallower quantum circuits. We assume that $M$ is an even number without loss of generality.
The method relies on the construction of $M$-order shadows from two $M/2$-order shadows. Based on the properties of the permutation group, we can express the calculation of the $M$-order function as
\begin{align}
    &\Tr(O_1\rho_e\dots O_M\rho_e)\nonumber \\= &\Tr(O_1\otimes\dots \otimes O_M S_{(1,M/2+1)}(\rho_{f,M/2}\otimes \rho_{f,M/2})),\label{eq:multi}
\end{align}
where $S_{(1,M/2+1)}$ is defined as
\begin{equation}
\begin{aligned}
   & S_{(1,M/2+1)} |\phi_1\rangle |\phi_2\rangle\dots|\phi_{M/2+1}\rangle \dots |\phi_M\rangle \\
   = & |\phi_{M/2+1}\rangle |\phi_2\rangle\dots|\phi_{1}\rangle\dots|\phi_M\rangle,
\end{aligned}
\end{equation}
and $\rho_{f,l}$ is the final state generated by the quantum circuit with depth $l - 1$ as shown in Fig.~\ref{fig: qubit efficient}.
Then, through Eq.~\eqref{eq:multi} we can get the corresponding unbiased estimator generated by the classical shadow,
\begin{equation}
\begin{aligned}
    &\Tr(O_1\rho_e\dots O_M\rho_e) \\= &\mathbf{E}_{i_1,i_2} \left[\Tr(\Tilde{O}_1\otimes \Tilde{O}_2 S_{(1,M/2+1)}\left(\mathcal{\widetilde{M}}(\hat{\rho}_{i_1})\otimes\mathcal{\widetilde{M}}(\hat{\rho}_{i_2})\right))\right],
\end{aligned}
\end{equation}
where $\Tilde{O}_1 = (X_0 + iY_0)\otimes O_1\otimes \dots \otimes O_{M/2}$ and  $\Tilde{O}_2 = (X_0 + iY_0)\otimes O_{M/2+1}\otimes \dots \otimes O_{M}$. The single-shot classical shadows $\mathcal{\widetilde{M}}(\hat{\rho}_{i_1})$ and $\mathcal{\widetilde{M}}(\hat{\rho}_{i_2})$ are $M/2$-order shadows, i,e, the shadows of $\rho_{f,M/2}$, and $\mathcal{\widetilde{M}}$ can be $\mathcal{\widetilde{M}}_P$, $\mathcal{\widetilde{M}}_C$ or $\mathcal{\widetilde{M}}_{\mathrm{mix}}$.
In addition, we can decompose $M$ into more parts to obtain results using shallower circuits. The formula for the general case is given by
\begin{equation}
\begin{aligned}
        &\Tr(O_1\rho_e\dots O_M\rho_e) \\=& \Tr(\Tilde{O}_1\otimes\dots \otimes \Tilde{O}_a \hat{S}(\rho_{f,M/a}\otimes\cdots \otimes \rho_{f,M/a}))\\=&\mathbf{E}_{i_1,\cdots,i_a}\left[\Tr(\Tilde{O}_1\otimes\dots \otimes \Tilde{O}_a \hat{S}(\mathcal{\widetilde{M}}(\hat{\rho}_{i_1})\otimes\cdots \otimes \mathcal{\widetilde{M}} (\hat{\rho}_{i_a})))\right],
\end{aligned}
\end{equation}
where $\hat{S} = S_{(1,(a-1)M/a+1)}S_{(1,(a-2)M/a+1)}\cdots S_{(1,M/a+1)}$, $\Tilde{O}_k = (X_0+iY_0)\otimes O_{M(k-1)/a+1}\otimes \cdots\otimes O_{Mk/a} $ and $a$ is the number of partitions of $M$. 
As a trade-off, the variance of the estimates with the given shots increases, which can be reduced by increasing the number of shots. The proposed method can outperform traditional state tomography methods. We present the specific results in the Theorem~\ref{theorem: 3}.

\section{Variance analysis}

\label{sec: scaling}

In this section, we analyze how the variance of the shadow estimation of an observable depends on the dimension $d=2^n$ of the state $\rho_e$ and the number of shots $N$. 
 The unbiased estimator of $\Tr\left(\tilde{O}\rho_f\right)$ is
\begin{equation}
    \hat{o}_i = \Tr(\Tilde{O}\mathcal{\widetilde{M}}(\hat{\rho}_i)),
\end{equation}
where $\hat{\rho}_i =  C_i^{\dagger}|z_i\rangle\langle z_i|C_i$.
Here $C_i$ and $|z_i\rangle$ are defined in Algorithm~\ref{alg:QESE}. The classical processing $\mathcal{\widetilde{M}}$ depends on the random unitary $C_i$ applied. Then we have the following theorem
\begin{theorem}
Suppose one shot is performed to estimate the $M$-order nonlinear function of a quantum state $\rho_e$ with dimension $d = 2^n$.
Then for Pauli-type shadow, the variance of estimator $\hat{o}_i$ is bounded by
\begin{align}
    \max(\mathbf{Var}(\mathbf{Re}(\hat{o}_i)),\mathbf{Var}(\mathbf{Im}(\hat{o}_i)))\leq 3\times 
 4^{\sum_{j=1}^M l_j} \|\Tilde{O}\|^2,
\end{align}
where $l_j$ is the locality of $O_j$ and $ \|\Tilde{O}\|^2 = \max_{1\leq j\leq M}\|O_j\|_{\infty}^2$.
\\

For Clifford-type shadow, the variance is bounded by
\begin{small}    
\begin{align}
    &\max(\mathbf{Var}(\mathbf{Re}(\hat{o}_i)),\mathbf{Var}(\mathbf{Im}(\hat{o}_i))) \nonumber \\\leq&
    3\left( \max_j\left(3\Tr(O_j^2) +\frac{2\Tr(O_j)\|O_{j,0}\|_{\infty} + \Tr(O_j)^2}{2^n}\right)\right)^{\mathcal{N}(\Tilde{O})-1},
\end{align}
\end{small}
where $\mathcal{N}(\Tilde{O})$ denotes how many operators in $\Tilde{O}$ are not identity operators. $O_{j,0} = O_j - \frac{\Tr(O_j)}{2^n}\text{I}_{2^n}$ is the traceless part of $O_j$.



\label{thm-1}
\end{theorem}
The proof is given in Appendix~\ref{proof: theorem1}.
This result shows a trade-off for Clifford-type shadows in qubit reduction, where the variance grows exponentially with the number of non-trivial observations $O_j$. In practical applications, $\mathcal{N}(\Tilde{O}) = 3$ is adequate for most cases~\cite{huang2020predicting,marrero2021entanglement,sack2022avoiding}. It also shows that the Clifford-type shadow here is between the Pauli-type and complete Clifford-type shadows. Its variance depends not only on the locality of a single Hermitian operator $O_j$ but also on the number of non-trivial operators in the joint operator $\Tilde{O}$.

\begin{figure}
\subfigure[ 
]
{\includegraphics[width=1\linewidth]{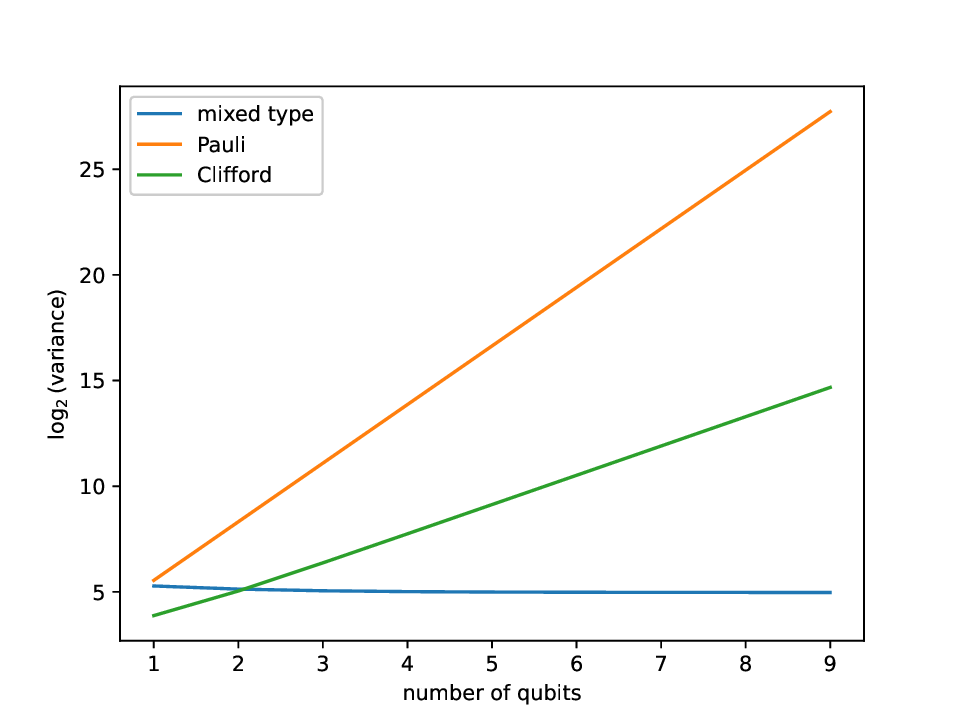}\label{fig:trade_off_qubits}}
\subfigure[
]{\includegraphics[width=1\linewidth]{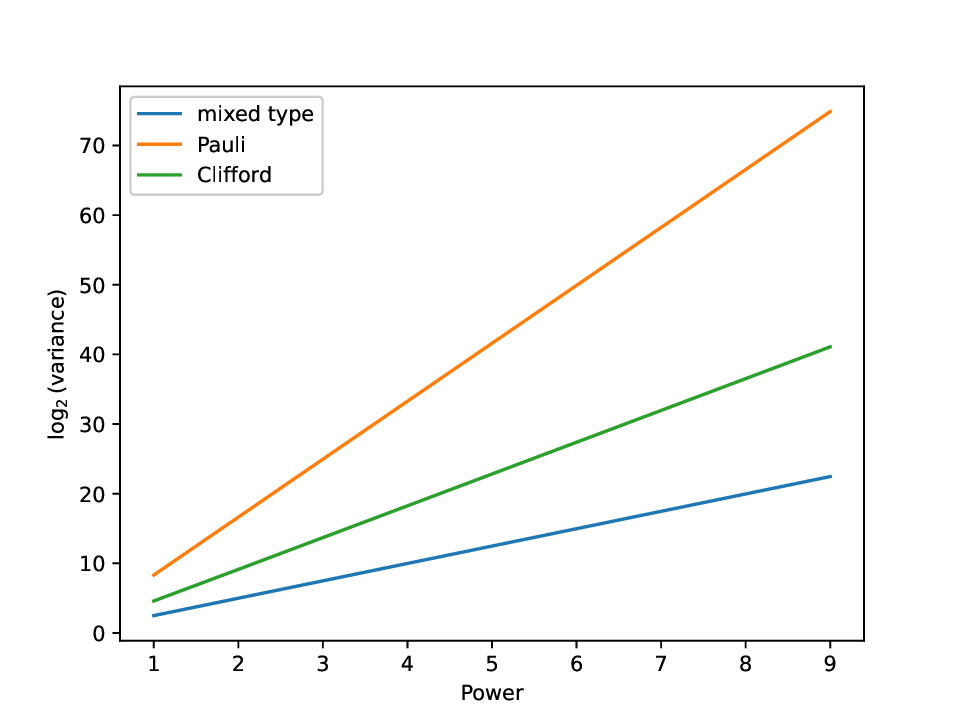}\label{fig:trade_off_power}}
\caption{Comparison of the variances of Pauli, Clifford, and mixed type shadows. (a) The variance of the estimation of $\Tr(O_1\rho_e O_2\rho_e)$ versus the number of qubits of $\rho_e$, where $O_1$ is local and $O_2$ is global. (b) The variance of the estimation of $\Tr\left((O_1\rho_e)^x (O_2\rho_e)^x\right)$, versus the order $x$, where we set $\rho_e$ to be a $5$-qubit state. The detailed simulation formula in given in Sec.~\ref{sec: num}. }
\end{figure}

In practical applications, operators $O_1,\dots,O_M$ may not be all local or global. Therefore, a mixed shadow approach, employing Pauli-type shadows for local observables and Clifford-type shadows for global observables, emerges as a better choice. The comparative analysis of the upper bound of the variance for Pauli-type, Clifford-type, and mixed-type shadows in the single-shot scenario is presented in Fig.~\ref{fig:trade_off_qubits} and Fig.~\ref{fig:trade_off_power}.



Next, we examine how the variance varies with the depth of the quantum circuit.
Let $M$ be the order of the virtual distillation that is assumed to be even. Suppose we use a circuit depth of $M/2-1$. Then the unbiased estimator is given by
\begin{equation}
\hat{o}_{i_1,i_2} = \Tr(\Tilde{O}_1\otimes \Tilde{O}_2 S_{(1,M/2+1)}(\tilde{\rho}_{i_1,i_2})),
\end{equation}
where
\begin{equation}
    \tilde{\rho}_{i_1,i_2}= \frac{1}{N(N-1)}\sum_{i_1\neq i_2}\mathcal{\widetilde{M}}(\hat{\rho}_{i_1})\otimes \mathcal{\widetilde{M}}(\hat{\rho}_{i_2}),
\end{equation}
where $i_1\neq i_2$ ensures independent shadows.
Then we have the following theorem.
\begin{theorem}
Suppose $N$ shots are performed to estimate the $M$-order nonlinear function of a quantum state $\rho_e$ with dimension $d=2^n$. The variance of the estimator $\hat{o}_{i_1,i_2}$ is bounded by
\begin{align}
    \max(\mathbf{Var}(\mathbf{Re}(\hat{o}_{i_1,i_2})),\mathbf{Var}(\mathbf{Im}(\hat{o}_{i_1,i_2}))) \leq
    \frac{p_1}{N} + \frac{p_2 d}{N} + \frac{p_3 d^{3}}{N^2},
\end{align}
for Pauli-type shadows, where $p_1, p_2, p_3$ are constants;
The variance is bounded by
\begin{align}
    &\max(\mathbf{Var}(\mathbf{Re}(\hat{o}_{i_1,i_2})),\mathbf{Var}(\mathbf{Im}(\hat{o}_{i_1,i_2}))) \nonumber\\ \leq&
    \frac{c_1}{N} + \frac{c_2 d}{N} + \frac{c_3 d}{N^2} + \frac{c_4 d^2}{N^2},
\end{align}
for Clifford-type shadows, where $c_1, c_2, c_3, c_4$ are constants.
\label{thm-2}
\end{theorem}
The proof is in Appendix~\ref{proof: mutiplicy}.
This theorem reveals that for Pauli-type shadows, the number of measurements needed to reach a given accuracy is $\mathcal{O}(d^{3/2})$ when $d$ is large. For Clifford-type shadows, the number of measurements is $\mathcal{O}(d)$.  

We can further generalize the result by using a circuit with depth $M/a-1$, where $a$ is a factor of $M$. The unbiased estimator is given by
\begin{equation}
\hat{o}_a = \Tr\left(\bigotimes_{k=1}^a \Tilde{O}_k \hat{S} \tilde{\rho}_a
   \right),
\end{equation}
where
\begin{align}
    \tilde{\rho}_a :=\frac{(N-a+1)!}{N!} \sum_{\{i_1,i_2,\dots,i_a\}}\bigotimes_{k=1}^a \mathcal{\widetilde{M}}(C_{i_k}^{\dagger}|z_{i_k}\rangle\langle z_{i_k}|C_{i_k}),
\end{align}
where the set $\{i_1,i_2,\dots,i_a\}$ only contains the elements where $i_1, i_2, \dots, i_a$ are not equal to each other. This ensures independent shadows.
The variance for Clifford-type shadow is given by the following theorem.
\begin{theorem}
    Suppose $N$ shots are performed to estimate the $M$-order nonlinear function of a quantum state $\rho_e$ with dimension $d=2^n$ using Clifford-type shadows. The variance of the estimator $\hat{o}_a$ is
    \begin{equation}
    \max(\mathbf{Var}(\mathbf{Re}(\hat{o}_a)),\mathbf{Var}(\mathbf{Im}(\hat{o}_a))) \sim \mathcal{O}((d/N)^a).
    \end{equation}
    \label{theorem: 3}
\end{theorem}
Please refer to the Appendix~\ref{proof: theorem 3} for the proof.
This theorem shows that for Clifford-type shadows when we use the shallower circuit to perform shadow estimation, the number of shots required grows linearly with $d$ to reach the same accuracy, which is a fundamental difference from Pauli-type shadows. It seems that the complexity of classical calculations depends on $N^a$. In fact, the complexity of classical calculations is far less than $N^a$, which is formulated by the following theorem.
\begin{theorem}
    Given the classical shadows $\tilde{\rho}$ constructed by $N$ snapshots, the classical computational complexity of calculating $\frac{(N-a+1)!}{N!}\sum_{i_1,\dots,i_a}\Tr(\left(\bigotimes_{k=1}^a \Tilde{O}_k\right)\hat{S}\left(\bigotimes_{k=1}^a\tilde{\rho}_{i_k}\right))$ can be $\sim\mathcal{O}(a^2N^3)$.
\end{theorem}
Please refer to the Appendix~\ref{classical algorithm}
for the concrete classical algorithm and complexity analysis.

\section{Numerical simulation}\label{sec: num}

\begin{figure*}
\subfigure[]{\includegraphics[width=0.4\linewidth]{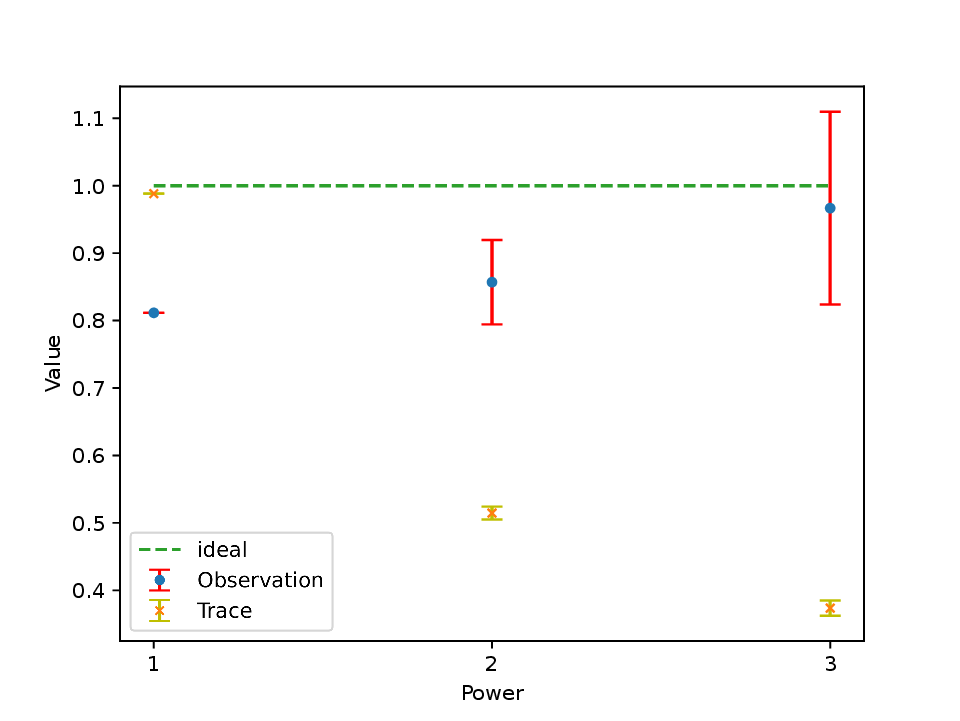}\label{fig:3GHZ_Pauli_01}}
\subfigure[]{\includegraphics[width=0.4\linewidth]{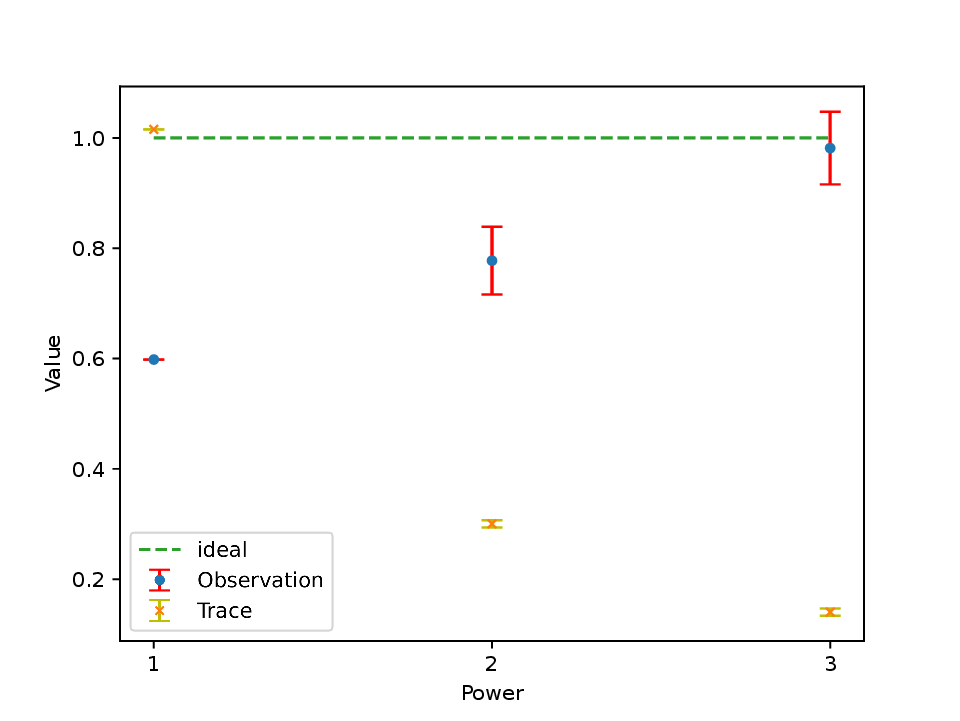}\label{fig:3GHZ_Pauli_02}}
\caption{Error-mitigated Pauli-type shadow estimations of a $3$-qubit GHZ state with different error rates. The horizontal axis is $\rho$ raised to the power. The ideal observation is $\langle\psi|O|\psi\rangle$, represented by the green dotted line, where $O = Z\otimes Z\otimes I$. The blue points are estimations of the observed quantities and the red lines are their error bars. The yellow points are estimations of $\Tr(\rho^M)$. Quantum circuits with lower error rates result in smaller errors and smaller error bars. (a) error rate $p=0.1$ and (b) error rate $p=0.2$.
 }

\subfigure[]{\includegraphics[width=0.4\linewidth]{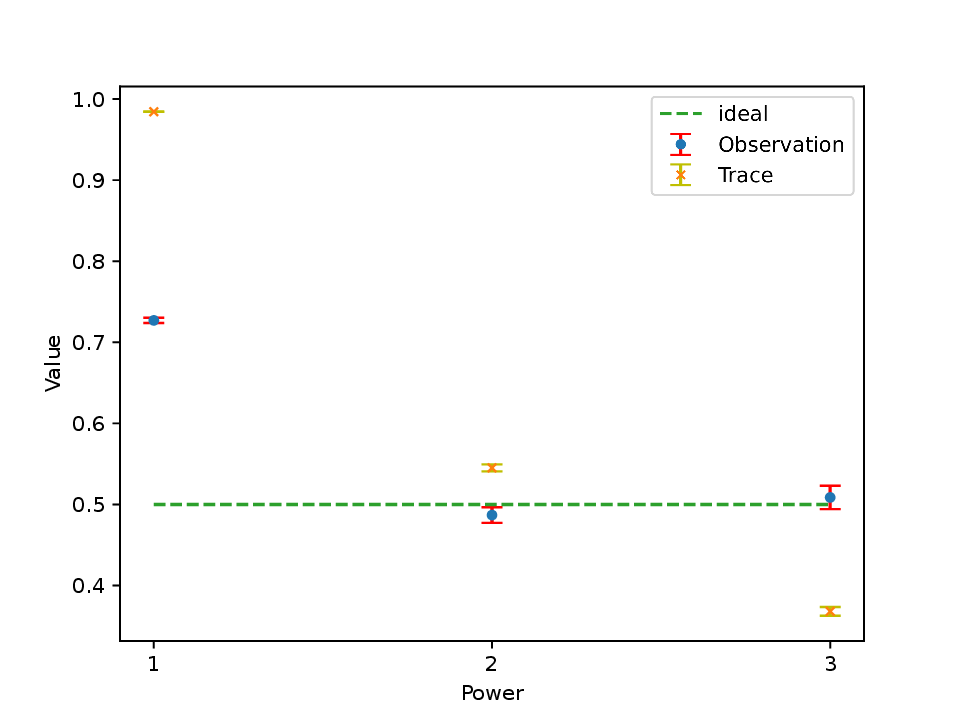}\label{fig:3GHZ_Clifford_01}}
\subfigure[]{\includegraphics[width=0.4\linewidth]{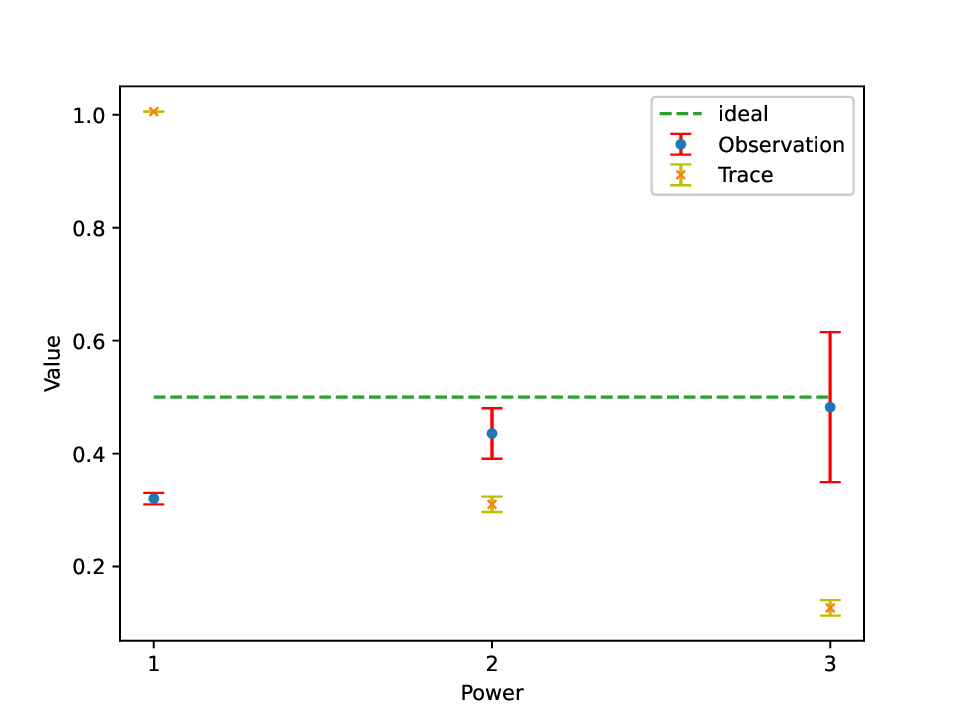}\label{fig:3GHZ_Clifford_02}}
    \caption{Error-mitigated Clifford-type shadow estimations a $3$-qubit GHZ state with different error rates. The horizontal axis is $\rho$ raised to the power. The ideal observation is $\langle\psi|O|\psi\rangle$, represented by the green dotted line, where $O = |0\rangle\langle 0|^{\otimes 3}$. The blue points are estimations of the observed quantities and the red lines are their error bars. The yellow points are estimations of $\Tr(\rho^M)$. Quantum circuits with lower error rates result in smaller errors and smaller error bars. (a) error rate $p=0.1$ and (b) error rate $p=0.2$.}

\end{figure*}

\begin{figure}
\subfigure[]{\includegraphics[width=1\linewidth]{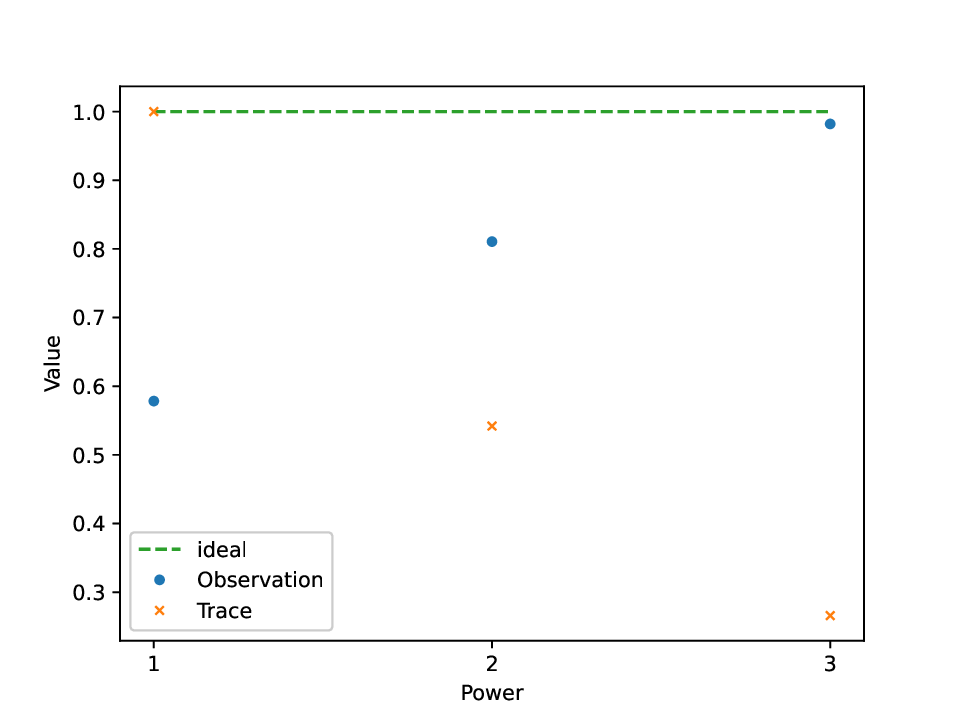}\label{fig:3GHZ_Pauli_two_01}}
\subfigure[]{\includegraphics[width=1\linewidth]{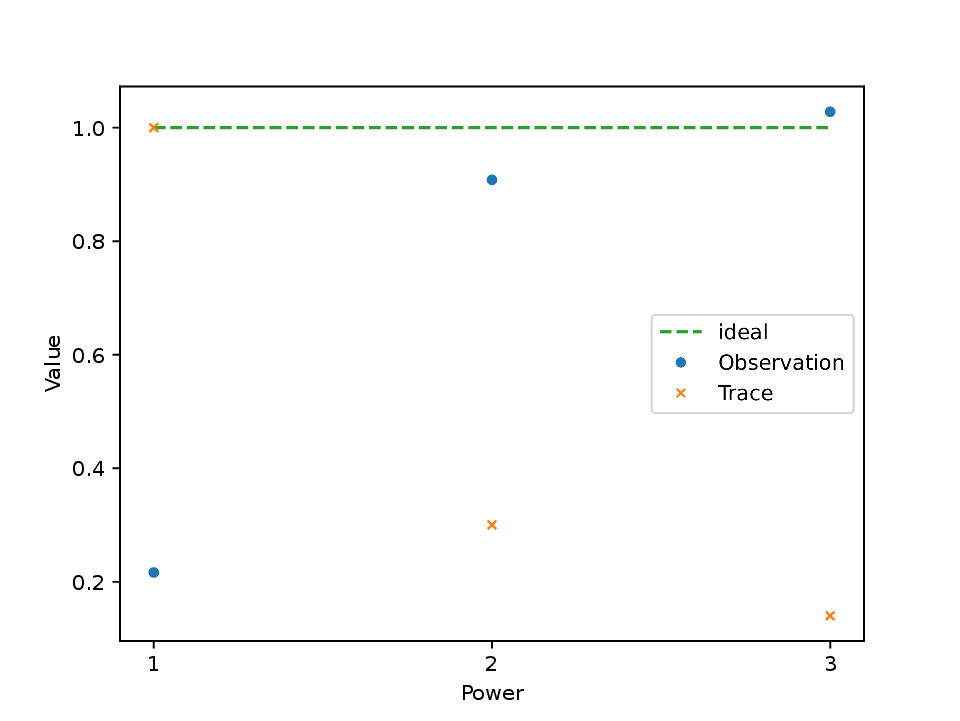}\label{fig:3GHZ_Pauli_two_02}}
    \caption{ 
    Error-mitigated Pauli-type shadow estimations for nonlinear function $\Tr(O\rho O\rho)$. The horizontal axis is $\rho$ raised to the power. The ideal observation is $\Tr(O\rho O\rho) = \langle\psi|O|\psi\rangle\langle\psi|O|\psi\rangle=1$, represented by the green dotted line, where $O = Z\otimes Z \otimes I$. The blue points are estimations of the observed quantities.  The yellow points are estimations of $\Tr(\rho_e^M)$ for different $M$. The observed value at $M=2$ surpasses the ideal expectation value due to the statistical fluctuations. (a) error rate $p=0.1$ and (b) error rate $p=0.2$.}
    \label{fig: extended VD}
\end{figure}

We first calculate the upper bound of the variance of the three different types of shadow estimations: Pauli-type, Clifford-type, and mixed-type. The variance for Pauli-type and Clifford-type shadow estimations can be bounded by Theorem~\ref{thm-1}. The variance of mixed-type shadow estimation is upper bound by
\begin{align}
        &\max(\mathbf{Var}(\mathbf{Re}(\hat{o}_i)),\mathbf{Var}(\mathbf{Im}(\hat{o}_i))) \leq 9\times\text{Var}_L\times\text{Var}_G.
\end{align}
where $\text{Var}_L = 4^{\sum_{j=1}^M l_j} \|\Tilde{O}_L\|^2$ is the contribution of local observables and 
\begin{small}
\begin{equation}
\begin{aligned}
&\text{Var}_G \\= &\left( \max_j\left(3\Tr(O_j^2) +2\Tr(O_j)\|O_{j,0}\|_{\infty} + \Tr(O_j)^2/{2^n}\right)\right)^{\mathcal{N}(\Tilde{O}_G)-1}
\end{aligned}
\end{equation}
\end{small}
comes from global ones. We use $\Tilde{O}_L$ and $\Tilde{O}_G$ to denote the local and global part of $\Tilde{O}$, respectively. For an observation in the form of $\Tr((O_1\rho_e)^x(O_2\rho_e))^x$, where $O_1$ is local and $O_2$ is global, it is straightforward that $\text{Var}_L = 4^{x}$ and $\text{Var}_G =  (3+(1/2)^n)^x$.

Next, we examine the effect of error mitigation.
We give two concrete examples of our algorithm calculating
\begin{align}
    \langle O\rangle = \frac{\Tr(O\rho_e^M)}{\Tr(\rho_e^M)},
\end{align}
and 
\begin{align}
    \langle O_1 O_2\rangle = \frac{\Tr(O_1\rho_e^{M} O_2\rho_e^{M})}{\Tr(\rho_e^{M})\Tr(\rho_e^{M})}.
\end{align}
In the numerical simulation, we set the ideal state to be a three-qubit GHZ state:
\begin{align}
    |\psi\rangle = \frac{1}{\sqrt{2}}(|000\rangle + |111\rangle).
\end{align}
We introduced a statistical noise $\Lambda$ in the following form 
\begin{align}
    \Lambda(\rho) = (1-p)\rho + pY\rho Y^{\dagger},
\end{align}
where $p$ is the error rate, and $Y$ denotes the Pauli-Y operator. In numerical simulations, we set the error rate $p$ to be $0.1$ and $0.2$.

In Fig.~\ref{fig:3GHZ_Pauli_01} and Fig.~\ref{fig:3GHZ_Pauli_02}, we show the impact of error mitigation on the classical shadow and the expected value of the linear observation. It can be found that as the order $M$ of VD increases, the estimated value of the linear observation quickly converges to the ideal value. At the same time, as a cost, $\Tr(\rho_e^M)$ significantly decreases. The standard deviation of the estimated value, i.e., the error bar in Fig.~\ref{fig:3GHZ_Clifford_01} and Fig.~\ref{fig:3GHZ_Clifford_02} increases with $M$.
Nonlinear observations exhibit similar behavior to linear observations, as shown in Fig.~\ref{fig: extended VD}.

\section{Conclusion and outlook}\label{sec: Con}
We introduce an error-mitigated shadow estimation method based on the VD and qubit reset, achieving robustness to gate errors. By analyzing the variance of shadow estimations and demonstrating the error mitigation effect through numerical simulations for linear and nonlinear functions, we showcase the strengths of our algorithm. Our approach utilizes qubit reset to reduce qubit overhead, enabling mixed Clifford and Pauli-type shadows. This unique feature offers a reduction in measurement requirements, particularly when dealing with predetermined observables. Furthermore, we explore the error-mitigated shadow estimation for a given accuracy using shallower quantum circuits and analyze the corresponding variances. It is worth noting that advanced techniques for classical shadow estimation or randomized measurements hold the potential to further improve our algorithm, either by reducing the complexity of quantum circuits or by shrinking the variance of the estimator, as highlighted in previous works~\cite{huang2021efficient,acharya2021shadow,zhang2023minimal,zhou2023performance,hu2023classical,hu2022hamiltonian,zhao2021fermionic,lukens2021bayesian,low2022classical,ippoliti2023operator,gu2023efficient,o2022fermionic,wan2023matchgate,bertoni2022shallow,shivam2023classical}. We leave the exploration of these potential combinations for future research.
\begin{acknowledgments}
This work is supported by National Natural Science Foundation of China Grants No. 12225507, 12088101, 61832003, 62272441, 12204489, and the Strategic Priority Research Program of Chinese Academy of Sciences Grant No. XDB28000000. 
\end{acknowledgments}


\bibliographystyle{IEEEtran}  
\bibliography{ref.bib}  

\onecolumngrid
\appendix
\section{Proof of Theorem \ref{thm-1}}
\label{proof: theorem1}
The variance for Pauli-type shadows can be directly derived from the Lemma 1 in Ref~\cite{huang2020predicting}. In this section, we focus on the variance for Clifford-type shadows. 
We use the notation $\hat{o}_{X_0} = \Tr(X_0\tilde{\rho})$ and $\hat{o}_{Y_0} = \Tr(Y_0\tilde{\rho})$. The variance of $\hat{o}_{Y_0}$ can be bounded by:
\begin{align}
    \mathbf{Var}[\hat{o}_{Y_0}] &\leq \max_{\sigma :\text{state}} \Tr(\sigma \sum_{z_{i}\in \{ 0,1   \}^{nM+1}} \mathbf{E}_{C_i \in \{ \mathrm{CI}^{n} \}^{\otimes M} }\left[C_i^{\dagger} |z_i\rangle\langle z_i|C_i \left|\langle z_i|C_i \mathcal{\widetilde{M}}_C(f_{Y_0})C_i^{\dagger}|z_i\rangle\right|^2\right]). 
\end{align}
To calculate the upper bound, we first calculate the average operator
\begin{equation}
\begin{aligned}
    & \sum_{z_i \in \{0,1  \}^{nM+1}} \mathbf{E}_{C_i \in \mathrm{CI}\otimes \{ \mathrm{CI}^{n} \}^{\otimes M} }  \left[C_i^{\dagger} |z_i\rangle\langle z_i|C_i \left|\langle z_i|C_i \mathcal{\widetilde{M}}_C(f_{Y_0})C_i^{\dagger}|z_i\rangle\right|^2\right] \nonumber\\
    & = 9\sum_{z_{0,i} \in\{0,1\}} \mathbf{E}_{C_{0,i}\in \mathrm{CI}}\left[C_{0,i}^{\dagger}|z_{0,i}\rangle\langle z_{0,i}| C_{0,i} \left|\langle z_{0,i}|C_{0,i}Y_0C_{0,i}^{\dagger}|z_{0,i}\rangle\right|^2 \right]\nonumber\\
    \quad& \otimes \sum_{z_{1,i}^A \in \{ 0,1 \}^{n}} \mathbf{E}_{C_{1,i}^A \in \mathrm{CI}^n} \left[C_{1,i}^{A\dagger}|z_{1,i}^A\rangle\langle z_{1,i}^A|C_{1,i}^A\left|\langle z_{1,i}^A C_{1,i}^A \mathcal{M}_C(O_1) C_{1,i}^{A\dagger} |z_{1,i}^A\rangle\right|^2\right]\nonumber\\ \quad&\bigotimes_{j=2}^M\sum_{z_{j,i}^B\in \{0,1\}^n}\mathbf{E}_{C_{j,i}^B \in \mathrm{CI}^n} \left[C_{j,i}^{B\dagger}|z_{j,i}^B\rangle\langle z_{j,i}^B|C_{j,i}^B\left|\langle z_{j,i}^B C_{j,i}^B \mathcal{M}_C(O_j) C_{j,i}^{B\dagger} |z_{j,i}^B\rangle\right|^2\right],
\end{aligned}
\end{equation}
according to the fact that 
\begin{equation}
    C_i = C_{M,i}^B\otimes C_{M-1,i}^B\otimes \dots \otimes C_{2,i}^B\otimes C_{1,i}^A\otimes C_{0,i},
\end{equation}
and 
\begin{equation}
    |z_i\rangle = |z_{M,i}^B\rangle \otimes |z_{M-1,i}^B\rangle\otimes \dots \otimes |z_{2,i}^B\rangle\otimes |z_{1,i}^A\rangle\otimes |z_{0,i}\rangle.
\end{equation}
Here $CI^n$ represents the Clifford group defined on $n$ qubits. The term corresponding to the ancilla qubit can be written as
\begin{equation}
\begin{aligned}
\Tilde{C}:= &\sum_{z_{0,i} \in \{0,1\}} \mathbf{E}_{C_{0,i}\in \mathrm{CI}}C_{0,i}^{\dagger}|z_{0,i}\rangle\langle z_{0,i}| C_{0,i} \left|\langle z_{0,i}|C_{0,i}Y_0C_{0,i}^{\dagger}|z_{0,i}\rangle\right|^2\nonumber\\
&= \frac{I_2}{3}.
\end{aligned}
\end{equation}
For simplicity, we define the traceless observable as 
\begin{align}
    O_{1,0} = O_1 - \frac{\Tr(O_1)}{2^n} I_{2^n}.
\end{align}
The linear map $\mathcal{M}_C$ for Clifford-type shadows can be represented as 
\begin{align}
    \mathcal{M}_C(O_i) 
    &= \mathcal{M}_C(O_{i,0}) + \frac{\Tr(O_i)}{2^n}I_{2^n}\nonumber \\
    &= (2^n+1)O_{i,0} + \frac{\Tr(O_i)}{2^n}I_{2_n}.
\end{align}
Next, we turn to calculate the average operator defined on register $A$.
\begin{align}
\Tilde{A}:=
    &\sum_{z_{1,i}^A \in \{ 0,1 \}^{n}} \mathbf{E}_{C_{1,i}^A \in \mathrm{CI}^n} C_{1,i}^{A\dagger}|z_{1,i}^A\rangle\langle z_{1,i}^A|C_{1,i}^A\left|\langle z_{1,i}^A C_{1,i}^A \mathcal{M}_C(O_1) C_{1,i}^{A\dagger} |z_{1,i}^A\rangle\right|^2\nonumber\\
    & = \sum_{z_{1,i}^A \in \{ 0,1 \}^{n}} \mathbf{E}_{C_{1,i}^A \in \mathrm{CI}^n}  C_{1,i}^{A\dagger}|z_{1,i}^A\rangle\langle z_{1,i}^A|C_{1,i}^A 
     ((2^n+1)^2 \left|\langle z_{1,i}^A C_{1,i}^A O_{1,0} C_{1,i}^{A\dagger} |z_{1,i}^A\rangle\right|^2 \nonumber\\
    &\quad+\frac{2(2^n+1)\Tr(O_1)}{2^n} \left(\langle z_{1,i}^A |C_{1,i}^A O_{1,0} C_{1,i}^{A\dagger}|z_{1,i}^A\rangle \right)+ \frac{\Tr(O_1)^2}{8^n}I_{2^n})\nonumber\\
    & = \sum_{z_{1,i}^A \in \{ 0,1 \}^{n}}\left(\frac{2^n+1}{(2^n+2)2^n}\left(\Tr(O_{1,0}^2) I_{2^n} + 2O_{1,0}^2\right) + \frac{2\Tr(O_1)}{4^n}O_{1,0} + \frac{\Tr(O_1)^2}{8^n}I_{2^n}\right)\nonumber\\
    & = \frac{2^n+1}{2^n+2}\left(\Tr(O_{1,0}^2) I_{2^n} + 2O_{1,0}^2\right) + \frac{2\Tr(O_1)}{2^n}O_{1,0} + \frac{\Tr(O_1)^2}{4^n}I_{2^n}.
\end{align}
Similarly, the average operator defined on register $B$ can be written as 
\begin{equation}
\begin{aligned}
    \tilde{B}:= &\bigotimes_{j=2}^M\sum_{z_{j,i}^B \in \{ 0,1 \}^{n}} \mathbf{E}_{C_{j,i}^B \in \mathrm{CI}^n} C_{j,i}^{B\dagger}|z_{j,i}^B\rangle\langle z_{j,i}^B|C_{j,i}^B\left|\langle z_{j,i}^B C_{j,i}^B \mathcal{M}_C(O_1) C_{j,i}^{B\dagger} |z_{j,i}^B\rangle\right|^2\nonumber\\
        & = \bigotimes_{j=2}^M\left(\frac{2^n+1}{2^n+2}\left(\Tr(O_{j,0}^2) I_{2^n} + 2O_{j,0}^2\right) + \frac{2\Tr(O_j)}{2^n}O_{j,0} + \frac{Tr(O_j)^2}{4^n}I_{2^n}\right).\\
\end{aligned}
\end{equation}
Thus the total variance can be bounded by
\begin{align}
       \mathbf{Var}(\hat{o}_{Y_0}) &\leq 
\max_{\sigma:state}\Tr\left(\sigma(\tilde{B}\otimes \tilde{A} \otimes \tilde{C})\right)\nonumber\\
    &\leq 3 \prod_{j=1}^M\left(3 \Tr(O_j^2) +\frac{2\Tr(O_j)||O_{j,0}||_{\infty} }{2^n} +\frac{\Tr(O_j)^2}{4^n}\right)\nonumber\\
  & \leq3\left( \max_j(3\Tr(O_j^2) +\frac{2\Tr(O_j)||O_{j,0}||_{\infty} }{2^n} + \frac{\Tr(O_j)^2}{4^n})\right)^\mathrm{\mathcal{N}(\Tilde{O}) - 1} .
\end{align}
The variance $\mathbf{Var}[\hat{o}_{X_0}]$ has the same upper bound with $\mathbf{Var}[\hat{o}_{Y_0}]$.


\section{Proof of Theorem \ref{thm-2}}
\label{proof: mutiplicy}
In this section, we prove Theorem~\ref{thm-2}. The proof for Pauli-type shadows can be found in Ref.~\cite{huang2020predicting}. Therefore we only give the proof for Clifford-type shadows.
The estimation can be obtained by
\begin{align}
    \Tr(O_1\rho O_2\rho_e \dots O_M\rho_e) = \mathbf{E}_{i_1,i_2}\left[\Tr\left( \Tilde{O}_1 \otimes\Tilde{O}_2 
    \left(S_{(1,M/2+1)}\left(\mathcal{\widetilde{M}}_{C}(\hat{\rho}_{i_1})\otimes \mathcal{\widetilde{M}}_{C}(\hat{\rho}_{i_2})\right)\right)\right)\right].
\end{align}
The swap operator $S_{(1,M/2+1)}$ can be expanded by Pauli operators defined on the corresponding qubits
\begin{align}
    S_{(1,M/2+1)} = \frac{1}{d}\Sigma_{s=1}^{d^2} P_s\otimes P_s,
\end{align}
where $d = 2^n$. Without loss of generality, suppose $\Tilde{O}_1 = X_0\otimes O_1\otimes \bigotimes_{j=2}^{M/2-1} I_{2}$ and $\Tilde{O}_2 = X_0\otimes O_2\otimes \bigotimes_{j=2}^{M/2-1} I_{2}$.
The estimation can be expanded as
\begin{align}
    \hat{o}_{X_0X_0}& =  \mathbf{E}_{i_1,i_2}\left[\Tr\left(\tilde{O}_1\otimes \tilde{O}_2 (S_{(1,M/2+1)}\left(\mathcal{\widetilde{M}}_{C}(\hat{\rho}_{i_1})\otimes \mathcal{\widetilde{M}}_{C}(\hat{\rho}_{i_2})\right)\right))\right]\nonumber\\
    & = \frac{1}{d}\sum_{s=1}^{d^2}  \mathbf{E}_{i_1,i_2}\left[\Tr\left(\mathcal{\widetilde{M}}_{C}(X_0\otimes O_1P_s) \hat{\rho}_{i_1}\right)  \Tr\left(\mathcal{\widetilde{M}}_{C}(X_0\otimes O_2P_s )\hat{\rho}_{i_2}\right)\right],
\end{align}
where we use $\hat{\rho}_{i_1}$ and $\hat{\rho}_{i_2}$ to denote the density matrices constructed from the single shot. Furthermore, the variance of the estimation $\hat{o}_{X_0X_0}$ is no more than
\begin{align}
    &\mathbf{E}_{i_1,i_2}(\hat{o}_{X_0X_0}^2) \nonumber\\=& \frac{1}{d^2} \mathbf{E}_{i_1,i_2}\left(\sum_{s=1}^{d^2} \Tr\left(\mathcal{\widetilde{M}}_{C}(X_0 O_1\otimes P_s) \hat{\rho}_{i_1}\right)  \Tr\left(\mathcal{\widetilde{M}}_{C}(X_0 \otimes O_2P_s) \hat{\rho}_{i_2}\right)\right)^2\nonumber\\
     = &\frac{1}{d^2}\mathbf{E}_{i_1,i_2}\left[\sum_{s_1=1}^{d^2}\sum_{s_2=1}^{d^2} \Tr\left(\mathcal{\widetilde{M}}_{C}(X_0\otimes O_1P_{s_1}) \hat{\rho}_{i_1}\right) \Tr\left(\mathcal{\widetilde{M}}_{C}(X_0 \otimes O_1P_{s_1}) \hat{\rho}_{i_2}\right)\Tr\left(\mathcal{\widetilde{M}}_{C}(X_0\otimes O_2 P_{s_2})\hat{\rho}_{i_1}\right)\Tr\left(\mathcal{\widetilde{M}}_{C}(X_0 \otimes O_2P_{s_2}) \hat{\rho}_{i_2}\right)\right].
\end{align}
Without loss of generality, we suppose that $\Tr({O}_1) = \Tr({O}_2) = 0$, both $O_1$ and $O_2$ are hermitian, and reversible. 
Note that the expectation $\mathbf{E}(\cdot)$ here means to take the expectation for a random Clifford gate before measurement and all possible measurement results. Because $i_1$ and $i_2$ are independent of each other, we can calculate the expectations separately. We need to discuss it on a case-by-case basis. For any pair of $P_{s_1}$ and $P_{s_2}$, there are four cases we need to discuss. If $O_1P_{s_1} = O_1P_{s_2} = I_{2^n}$, then
\begin{align} 
&\mathbf{E}_{i_1}
\Tr(\mathcal{\widetilde{M}}_{C}(X_0\otimes O_1P_{s_1})\hat{\rho}_{i_1})\Tr(\mathcal{\widetilde{M}}_{C}(X_0\otimes O_1P_{s_2})\hat{\rho}_{i_1})\nonumber\\
=& 9\mathbf{E}_{i_1}\Tr(X_0\mathcal{{M}}_{C}(O_1P_{s_1})\hat{\rho}_{i_1})\Tr(X_0\mathcal{{M}}_{C}(O_1P_{s_2})\hat{\rho}_{i_1})\nonumber\\
=& 9\mathbf{E}_{i_1}\Tr(X_0\hat{\rho}_{i_1})\Tr(X_0\hat{\rho}_{i_1})\nonumber\\
\leq & 27.
\end{align}
After traversing $P_{s_1}$ and $P_{s_2}$, only one term satisfies $O_1P_{s_1} = O_1P_{s_2} = I_{2^n}$.
If $O_1P_{s_1} = O_1P_{s_2} \neq I_{2^n}$, then
\begin{align} 
&\mathbf{E}_{i_1}
\Tr(\mathcal{\widetilde{M}}_{C}(X_0\otimes O_1P_{s_1})\hat{\rho}_{i_1})\Tr(\mathcal{\widetilde{M}}_{C}(X_0\otimes O_1P_{s_2})\hat{\rho}_{i_1})\nonumber\\
=& 9\mathbf{E}_{i_1}\Tr(X_0\mathcal{{M}}_{C}(O_1P_{s_1})\hat{\rho}_{i_1})\Tr(X_0\mathcal{{M}}_{C}(O_1P_{s_2})\hat{\rho}_{i_1})\nonumber\\
=& 9(d+1)^2\mathbf{E}_{i_1}\Tr((X_0\otimes O_1P_{s_1})\hat{\rho}_{i_1})\Tr((X_0\otimes O_1P_{s_2})\hat{\rho}_{i_1})\nonumber\\
\leq & 27(d+1)^2\max_{\sigma,state}\Tr(\sigma\frac{\Tr( O_1P_{s_1}O_1P_{s_2})I_{2^n} + O_1P_{s_1}O_1 P_{s_2} + O_1P_{s_2}O_1P_{s_1}}{(d+2)(d+1)})\nonumber\\
= &27(d+1)
\end{align}
There are totally $(d^2-1)$ terms satisfying the condition $O_1P_{s_1} = O_1P_{s_2} \neq I_{2^n}$.
If $P_{s_1} \neq P_{s_2} = I_{2^n}$, then
\begin{align} 
&\mathbf{E}_{i_1}
\Tr(\mathcal{\widetilde{M}}_{C}(X_0\otimes O_1P_{s_1})\hat{\rho}_{i_1})\Tr(\mathcal{\widetilde{M}}_{C}(X_0\otimes O_1P_{s_2})\hat{\rho}_{i_1})\nonumber\\
=& 9(d+1)\mathbf{E}_{i_1}\Tr((X_0\otimes O_1P_{s_1})\hat{\rho}_{i_1})\Tr((X_0\otimes O_1P_{s_2})\hat{\rho}_{i_1})\nonumber\\
\leq & 27(d+1)\max_{\sigma,state}\Tr(\sigma\frac{O_1P_{s_1}}{(d+1)})\nonumber\\
= &27
\end{align}
There are totally $2(d^2-1)$ terms satisfying the condition.
If $O_1P_{s_1} \neq O_1P_{s_2}, O_1P_{s_1}\neq O_1I_{2^n} , O_1P_{s_2}\neq O_1I_{2^n}$, then

\begin{align} 
&\mathbf{E}_{i_1}
\Tr(\mathcal{\widetilde{M}}_{C}(X_0\otimes O_1P_{s_1})\hat{\rho}_{i_1})\Tr(\mathcal{\widetilde{M}}_{C}(X_0\otimes O_1P_{s_2})\hat{\rho}_{i_1})\nonumber\\
=& 9(d+1)^2\mathbf{E}_{i_1}\Tr((X_0\otimes O_1P_{s_1})\hat{\rho}_{i_1})\Tr((X_0\otimes O_1P_{s_2})\hat{\rho}_{i_1})\nonumber\\
\leq & 27(d+1)^2\max_{\sigma,state}\Tr(\sigma\frac{\Tr( O_1P_{s_1}O_1P_{s_2})I_{2^n} + O_1P_{s_1} O_1P_{s_2} + O_1P_{s_2}O_1P_{s_1}}{(d+2)(d+1)})\nonumber\\
= &54\frac{d+1}{d+2}
\end{align}
No more than $d^3$ terms are satisfying the condition.
As a summary, the variance for one shot is bounded by
\begin{align}
\mathbf{Var}[\hat{o}_{X_0 X_0}]\leq 27^2 \times \frac{1+(d+1)^2(d^2-1)+2(d^2-1)+4(d^2-1)^2}{d^2} \sim O(d^2).
\end{align}
The estimation given $N$ shots can be derived by
\begin{align}
    \hat{o}_{X_0X_0}& = \frac{1}{N(N-1)}\sum_{\{i_1\neq i_2\}}\Tr\left(X_0\otimes X_0 S_{1,M_1+1}(\widetilde{\mathcal{M}}_C(\hat{\rho}_{i_1})\otimes \widetilde{\mathcal{M}}_C(\hat{\rho}_{i_2}))\right)\nonumber\\
    & = \frac{1}{dN(N-1)}\sum_{\{i_1\neq i_2\}}\sum_{s=1}^{d^2} \Tr(X_0 \mathcal{M}_{C}(O_1P_s) \hat{\rho}_{i_1})  \Tr(X_0 \mathcal{M}_{C}(O_2P_s) \hat{\rho}_{i_2}),
\end{align}
The variance for $N$ shots can be bounded by
\begin{align}
    \mathbf{E}(\hat{o}_{X_0 X_0}^2) &= \frac{1}{d^2N^2(N-1)^2} \sum_{\{i_1\neq i_2\}}\sum_{\{i_1^{\prime}\neq i_2^{\prime}\}}\mathbf{E}\left(\sum_{s=1}^{d^2} \Tr\left(X_0 \mathcal{{M}}_{C}(O_1P_s) \hat{\rho}_{i_1}\right)  \Tr\left(X_0 \mathcal{M}_{C}(O_2P_s) \hat{\rho}_{i_2}\right)\right)^2\nonumber\\
    & = \frac{1}{d^2N^2(N-1)^2} \sum_{\{i_1\neq i_2\}}\sum_{\{i_1^{\prime}\neq i_2^{\prime}\}}\mathbf{E}\sum_{s_1=1}^{d^2}\sum_{s_2=1}^{d^2} \Tr\left(X_0 \mathcal{{M}}_{C}(O_1P_{s_1}) \hat{\rho}_{i_1}\right))\nonumber\\  
    &\times \left( \Tr\left(X_0 \mathcal{{M}}_{C}(O_1P_{s_2}) \hat{\rho}_{i_1^{\prime}}\right)\Tr\left(X_0 \mathcal{{M}}_{C}(O_2P_{s_1})\hat{\rho}_{i_2}\right)\Tr\left(X_0 \mathcal{{M}}_{C}(O_2P_{s_2}) \hat{\rho}_{i_2^{\prime}}\right)\right).
\end{align}
We need to discuss the situation again. There are only two following cases to consider since other cases do not contribute to the variance.
The first case we need to discuss is where $i_1^{\prime}= i_1$ and $i_2^{\prime}= i_2$. In this case, the corresponding variance is
\begin{align}
&\quad \frac{N(N-1)}{d^2N^2(N-1)^2} \mathbf{E}\left(\sum_{s=1}^{d^2} \Tr\left(X_0 \mathcal{{M}}_{C}(O_1P_s) \hat{\rho}_{i_1}\right)  \Tr\left(X_0 \mathcal{{M}}_{C}(O_2P_s) \hat{\rho}_{i_2}\right)\right)^2\nonumber\\
    & = \frac{1}{d^2N(N-1)} \mathbf{E}\left(\sum_{s_1=1}^{d^2}\sum_{s_2=1}^{d^2} \Tr\left(X_0 \mathcal{{M}}_{C}(O_1P_{s_1}) \hat{\rho}_{i_1}\right)  \Tr\left(X_0 \mathcal{{M}}_{C}(O_1P_{s_2}) \hat{\rho}_{i_1}\right)\Tr\left(X_0 \mathcal{{M}}_{C}(O_2P_{s_1})\hat{\rho}_{i_2}\right)\Tr\left(X_0 \mathcal{{M}}_{C}(O_2P_{s_2}) \hat{\rho}_{i_2}\right)\right)\nonumber\\
    &\leq \frac{1  + 2(d-1)(d^2-1) + (d^2-1)^2 (d+1)/(d+2)}{d^2N(N-1)}\nonumber\\
    &\sim\mathcal{O}\left(\frac{d^2}{N^2}\right).
\end{align}
For the case where $i_1^{\prime} = i_1$ and $i_2^{\prime}\neq i_2$, the variance is 
\begin{align}
&\quad \frac{N(N-1)^2}{d^2N^2(N-1)^2} \mathbf{E}\left(\sum_{s_1=1}^{d^2}\sum_{s_2=1}^{d^2} \Tr\left(X_0 \mathcal{{M}}_{C}(O_1P_{s_1}) \hat{\rho}_{i_1}\right)\times \Tr\left(X_0 \mathcal{{M}}_{C}(O_1P_{s_2}) \hat{\rho}_{i_1}\right)\Tr\left(X_0 \mathcal{{M}}_{C}(O_2P_{s_1})\hat{\rho}_{i_2}\right)\Tr\left(X_0 \mathcal{{M}}_{C}(O_2P_{s_2}) \hat{\rho}_{i_2^{\prime}}\right)\right)\nonumber\\
&\leq \frac{3^5 \times (d+5d^2-3)}{d^2 N}\nonumber\\
&\sim \mathcal{O}\left(\frac{1}{N}\right).
\end{align}
Therefore the upper bound of the variance $\mathbf{Var}[\hat{o}_{X_0X_0}]$ is $\sim\mathcal{O}\left(\frac{d^2}{N^2}\right)$.

\section{Proof of Theorem \ref{theorem: 3}}
\label{proof: theorem 3}
Without loss of generality, suppose $\Tilde{O}_k = X_0\otimes O_k \otimes \bigotimes_{j=2}^{M/a-1} I_{2}$ where $k=1,2,\dots,a$.
Similarly to the last section, we first consider the case of single-shot, where the estimation is denoted by $\hat{o}$. 
\begin{align}
    \mathbf{E}[\hat{o}^2] &= \frac{1}{d^{2a}}\mathbf{E}\left(\sum_{s_1=1}^{d^2}\cdots \sum_{s_{a-1}=1}^{d^2} \Tr( 
    (
    X_0\otimes O_1P_{s_1}P_{s_2}\cdots P_{s_{a-1}}
    )\otimes
    (X_0\otimes O_2P_{s_1})\otimes\cdots (X_0\otimes O_{a}P_{s_{a-1}})
    \times \tilde{\rho}_{i_1}\otimes \tilde{\rho}_{i_2}\otimes \cdots\otimes \tilde{\rho}_{i_{a}})
    \right)^2\nonumber\\
    & =\frac{1}{d^{2a}}\sum_{s_1=1}^{d^2}\sum_{s_1 ^{\prime}=1}^{d^2}\cdots \sum_{s_{a-1}=1}^{d^2}\sum_{s_{a-1} ^{\prime}=1}^{d^2}\mathbf{E}\left[\Tr\left((X_0\otimes O_1P_{s_1}P_{s_2}\cdots P_{s_{a-1}})\tilde{\rho}_{i_1}
    \right)\Tr\left((X_0\otimes O_1P_{s_1^{\prime}}P_{s_2^{\prime}}\cdots P_{s_{a-1}^{\prime}})\tilde{\rho}_{i_1}\right)\right]
    \nonumber\\
    \quad&\times \mathbf{E}\left[\Tr\left((X_0\otimes O_2P_{s_1})  \tilde{\rho}_{i_2} \right)
    \Tr\left((X_0\otimes O_2P_{s_1^{\prime}})  \tilde{\rho}_{i_2} \right)\right]\times\cdots\times 
    \mathbf{E}\left[\Tr\left((X_0\otimes O_{a}P_{s_{a-1}})  \tilde{\rho}_{i_{a}} \right)
    \Tr\left((X_0\otimes O_{a}P_{s_{a-1}^{\prime}})  \tilde{\rho}_{i_{a}} \right)\right].
\end{align}
We know from the last section that
\begin{align}
    \mathbf{Var}[\hat{o}] \sim O(d^a)
\end{align}
Now we turn to multiple shots:
\begin{align}
    \mathbf{E}[\hat{o}_a^2] &= \frac{1}{d^{2a}N^{2a}} \sum\mathbf{E}\left[\Tr\left((X_0\otimes {O}_1P_{s_1}P_{s_2}\cdots P_{s_{n-1}})\tilde{\rho}_{i_1}
    \right)\Tr\left((X_0\otimes {O}_1P_{s_1^{\prime}}P_{s_2^{\prime}}\cdots P_{s_{a-1}^{\prime}})\tilde{\rho}_{i_1^{\prime}}
    \right)\right]\nonumber\\
    &\times \mathbf{E}\left[\Tr\left((X_0\otimes {O}_2 P_{s_1})  \tilde{\rho}_{i_2} \right)
    \Tr\left((X_0\otimes {O}_2P_{s_1^{\prime}})  \tilde{\rho}_{i_2^{\prime}} \right)\right]\cdots 
    \mathbf{E}\left[\Tr\left((X_0\otimes {O}_{a}P_{s_{a-1}})  \tilde{\rho}_{i_{a}} \right)
    \Tr\left((X_0\otimes {O}_{a}P_{s_{a-1}^{\prime}})  \tilde{\rho}_{i_{a}^{\prime}} \right)\right].
\end{align}
The largest term appears at $i_1=i_1^{\prime}, i_2=i_2^{\prime},\dots,i_{a}=i_{a}^{\prime}$. There are totally $N^a$ terms. Then we have
\begin{align}
    \mathbf{E}[\hat{o}^2]\sim O\left(\frac{d^a}{N^a}\right).
\end{align}
\section{Classical Algorithm for Calculating $\Tr(\bigotimes_{k=1}^a \tilde{O}_k\hat{S}\tilde{\rho}_a)$ and Complexity Analysis}
\label{classical algorithm}

In this section, we delve into the details of a specific classical algorithm and evaluate its computational complexity.  We use $\tilde{\rho}$ to represent the concrete classical shadows. Specifically: \begin{equation}
    \tilde{\rho} = \frac{1}{N}\sum_{i=1}^N \tilde{\rho}_{i} = \frac{1}{N} \sum_{i=1}^N \widetilde{M}(\hat{\rho}_i) = \frac{1}{N}\sum_{i=1}^N \tilde{\rho}_{i}^0\otimes\left(
(d+1)\hat{\rho}_{i}^{(1)}-I_{2^n}
\right)\otimes\left(\bigotimes_{j=2}^{M/a}\Tilde{\rho}_i^{(j)}
\right). 
\end{equation}

We have defined $\X_{i} =C_{i}^{\dagger}|z_{i}\rangle\langle z_{i}|C_{i}$, $\tilde{\rho}_{i}^0 = 3\hat{\rho}_i^0 - I_2$ and $\tilde{\rho}_i^{(j)}=(d+1)\hat{\rho}_i^{(j)} - I_{2^n}$. Here $\tilde{\rho}_i^0$ is the classical shadow of the ancilla qubit, and $\tilde{\rho}_i^{(j)}$ is the classical shadow of $j$-th $n$-qubit subsystem. $\hat{\rho}_{i}^{(1)}$ is the density operator defined on the $1$-st $n$-qubit system.
Recall that \begin{equation}
\Tilde{\rho}_a = \frac{(N+a-1)!}{N!} \sum_{\{i_1,i_2,\dots,i_a\}}\bigotimes_{k=1}^a\Tilde{\rho}_{i_k}.
\end{equation}

 We decompose the single shot classical shadow $\Tilde{\rho}_i$ into two parts: $\Tilde{\rho}_i = \Tilde{\rho}_{i,0} - \Tilde{\rho}_{i,1}$, where $\Tilde{\rho}_{i,0} = \Tilde{\rho}_{i}^0\otimes(d+1)\hat{\rho}_i^{(1)}\otimes \left(\bigotimes_{j=2}^{M/a}\Tilde{\rho}_{i}^{(j)} \right) $ 
 and $\Tilde{\rho}_{i,1} = \Tilde{\rho}_{i}^0\otimes I_{2^n} \otimes \left(\bigotimes_{j=2}^{M/a}\Tilde{\rho}_i^{(j)}\right)$.
Furthermore, the tensor product $\bigotimes_{k=1}^a\Tilde{\rho}_{i_k}$ can be broken down into $2^a$ terms as
\begin{equation}
   \bigotimes_{k=1}^a \Tilde{\rho}_{i_k} = \sum_{r_1,\dots,r_a}\bigotimes_{k=1}^a\Tilde{\rho}_{i_k,r_k},
   \end{equation}
where each $r_k$ is either $0$ or $1$.
We decompose $\bigotimes_{k=1}^a \Tilde{\rho}_{i_k}$ into a summation of $2^a$ terms because it facilitates the management of the swap operator $\hat{S}$, simplifying the overall process. We will see at the end of this section that we do not need to calculate each term separately and then sum, thus avoiding the exponential dependence on $a$. To gain a comprehensive understanding of the entire algorithm, we first focus on the scenario where all $r_k=0$ for $k=1,2,\dots,a$.
Then the effect of $\hat{S}$ can be written as
\begin{equation}
\hat{S}\bigotimes_{k=1}^a\Tilde{\rho}_{i_k,0} =\sum_{\{i_1,\dots,i_a\}}D_{i_a,i_1} \otimes D_{i_1,i_2}\otimes D_{i_2,i_3}\otimes \cdots \otimes D_{i_{a-1},i_a},
\end{equation}
where we have defined
\begin{equation}
\begin{aligned}
D_{x,y} &:=  \Tilde{\rho}_{y}^0\otimes \left(
(d+1)C_x^{\dagger}|z_x\rangle\langle z_y| C_y
\right) \otimes \left(
\bigotimes_{j=2}^{M/a}  \Tilde{\rho}_{y}^{(j)}  
\right).
\end{aligned}
\end{equation}
What we need to calculate is actually
\begin{equation}
\Tr(\left(\bigotimes_{k=1}^a \Tilde{O}_k\right)
\hat{S}\left( \bigotimes_{k=1}^a\Tilde{\rho}_{i_k,0}
\right)
) = \sum_{\{i_1,\dots,i_a\}}\Tr(\Tilde{O}_1 D_{i_a,i_1})\times\Tr(\Tilde{O}_2 D_{i_1,i_2})\times \dots \times \Tr(\Tilde{O}_aD_{i_{a-1},i_a}).
\label{eq: final calculation}
\end{equation}
Define the matrices $E_k$ with elements $(E_k)_{x,y} = \Tr(O_k D_{x,y})$ and $(E_k)_{x,y}=0$ when $x=y$.
Eq.~\eqref{eq: final calculation} can be interpreted as a matrix multiplication followed by a trace operation, specifically:
 \begin{equation}
 \Tr(\left(\bigotimes_{k=1}^a \Tilde{O}_k\right)
\hat{S}\left( \bigotimes_{k=1}^a\Tilde{\rho}_{i_k,0}
\right)
) = \Tr(
\prod_{k=1}^a E_k
)
\end{equation}

Next, we turn our attention to the remaining $2^a-1$ terms. Initially, we investigate the scenario where $r_{k_1}=1$, while all other $r_k$ values are set to $0$. In this case, we obtain. 
\begin{align}
\Tr( \left(\bigotimes_{k=1}^a\Tilde{O}_k\right) \hat{S} \left(\bigotimes_{k=1}^{k_1-1} \Tilde{\rho}_{i_k,0}\right) \otimes \Tilde{\rho}_{i_{k_1},1} \otimes \left(
\bigotimes_{k=k_1+1}^a\Tilde{\rho}_{i_k,0}
\right)
) = \Tr(\left(\prod_{k=1}^{k_1-1}E_{k}\right) E_{k_1,k_1+1} \left(\prod_{k=k_1+2}^a E_k\right) ),
\end{align}
where the elements of matrix $E_{k_1,k_1+1}$ are defined as $$
\begin{aligned}
(E_{k_1,k_1+1})_{x,y} = &\Tr(O_{0}\Tilde{\rho}_{x+1}^0)
\Tr(O_{0}\Tilde{\rho}_{y}^0)\Tr(O_{M(k_1-1)/a+1}O_{Mk_1/a+1}C_x^{\dagger}|z_x\rangle\langle z_y|C_y)\\
&\times\prod_{j=2}^{M/a}\Tr(O_{M(k_1-1)/a+j}\Tilde{\rho}_{x+1}^{(j)})\Tr(O_{Mk_1/a+1}\Tilde{\rho}_{y}^{(j)}),
\end{aligned}
$$
Here we have defined $O_0 = X_0 + iY_0$.
Extending this concept further, we introduce the matrix $E_{k_1,k_1^{\prime}}$ for the case where $k_1<k_1^{\prime}$, with its elements given by
\begin{equation}
\begin{aligned}
(E_{k_1,k_1^{\prime}})_{x,y} =&
\Tr(O_0\Tilde{\rho}_{x+1}^0)
\Tr(O_0\Tilde{\rho}_{y}^0)\Tr(O_{M(k_1-1)/a+1}O_{M(k_1^{\prime}-1)/a+1}C_x^{\dagger}|z_x\rangle\langle z_y|C_y)
\\
&\times\prod_{j=2}^{M/a}\Tr(O_{M(k_1-1)/a+j}\Tilde{\rho}_{x+1}^{(j)})\Tr(O_{M(k_1^{\prime}-1)/a+j}\Tilde{\rho}_{y}^{(j)})\\
&\times \prod_{k=k_1+1}^{k_1^{\prime}-1} \sum_{i_k=1}^N\Tr(O_0\Tilde{\rho}_{i_k}^0)\left(\prod_{j=2}^{M/a}\Tr(O_{M(k-1)/a+j}\Tilde{\rho}_{i_k}^{(j)})
\right).
\end{aligned}
\end{equation}
The computation of a single matrix element in $E_{k_1,k_1^{\prime}}$ has a complexity $\sim\mathcal{O}(|k_1^{\prime}-k_1|N/a)$. Given that $|k_1^{\prime}-k_1|$ is less than $a$ this complexity simplifies to $\sim\mathcal{O}(N)$.
 Considering the totality of matrices, there are $a(a-1)/2$  such matrices, each containing $N^2$ elements. Therefore, the overall complexity for computing all matrices $E_{k_1,k_1^{\prime}}$ with $1\leq k_1<k_1^{\prime}\leq a$ amounts to $\sim\mathcal{O}(a^2N^3)$.

To simplify notation, we define $E_{k,k} = E_k$ for $k=1,2,\dots,a$.
Ultimately, our goal is to compute the following expression:
\begin{equation}
    \begin{aligned}
        \sum_{i_1,\dots,i_a}\Tr\left( \left(\bigotimes_{k=1}^a \Tilde{O}_k 
        \right)
        \hat{S}\left(
        \bigotimes_{k=1}^a\Tilde{\rho}_{i_k}
        \right)
        \right) =
        \sum_{\substack{1 = k_1\leq k_1^{\prime}   <\cdots < k_b\leq k_b^{\prime}= a,\\ b\leq a}}
        \Tr(E_{k_1,k_1^{\prime}} E_{k_2,k_2^{\prime}} 
        \cdots E_{k_b,k_b^{\prime}}).
        \label{eq: without S}
    \end{aligned}
\end{equation}
So far, we have fully integrated the influence of $\hat{S}$ into a series of matrices $E_{k,k^{\prime}}$'s. However, currently, the right-hand side of Eq.\eqref{eq: without S} still comprises the summation of $2^a$ terms.
The naive complexity for summing over all valid combinations is  $\sim\mathcal{O}(2^a)$. Designing tailored algorithms can effectively mitigate computational complexity. 
For the sake of clarity in describing our approach, let us define
\begin{equation}
F^{(\Bar{k})} = \sum_{\substack{1= k_1\leq k_1^{\prime} <\cdots < k_{b}\leq k_{b}^{\prime}= \Bar{k},\\ b\leq \Bar{k}}} E_{k_1,k_1^{\prime}} \cdots E_{k_{b},k_{b}^{\prime}} ,
\end{equation}
for $\Bar{k}\geq 2$.
Assuming the complexity of computing the matrices $F^{(2)}, F^{(3)},\dots, F^{(\Bar{k})}$ is totally denoted by $G(\Bar{k})$ and given that we have already computed $F^{(2)}, F^{(3)},\dots, F^{(\Bar{k})}$, we can derive $F^{(\Bar{k}+1)} = \sum_{k=2}^{\Bar{k}}\Tr( F^{(k)}E_{k,\Bar{k}+1}) + E_{1,\Bar{k}+1}$. Consequently, the computational complexity of calculating only $F^{(\Bar{k}+1)}$ is $\sim\mathcal{O}(\Bar{k}N^3)$. An efficient algorithm involves sequentially computing $F^{(2)}, F^{(3)},\dots,F^{(a-1)}$, culminating in the determination of the matrix $F^{(a)}$. Note that $G(2)\sim\mathcal{O}(N^3)$.
Therefore, the overall complexity for computing $F^{(a)}$ is given by $G(a) \sim\mathcal{O}(a^2N^3)$ given all matrices $E_{k,k^{\prime}}$ are available. Since the complexity of computing all these matrices is also $\sim\mathcal{O}(a^2N^3)$, the total complexity for obtaining $F^{a}$ remains at  $\sim\mathcal{O}(a^2N^3)$. Indeed, when $\Tilde{O}_k$ assumes specific forms, such as $\Tilde{O}_1=O^0\otimes\bigotimes_{j=1}^{M/a}O_{j}$ and the remaining $\Tilde{O}_k=O^0\otimes I_{2}^{\otimes M/a}$, the computational complexity can be further optimized. However, we will not delve into the details here.

\end{document}